\documentclass[lettersize,journal]{IEEEtran}
\usepackage[T1]{fontenc}
\input{glyphtounicode}
\usepackage{amsmath,amsfonts}
\usepackage{algorithmic}
\usepackage{epsfig, algorithm}
\usepackage{array}
\usepackage{amssymb}
\usepackage{acronym} 
\usepackage[caption=false,font=normalsize,labelfont=sf,textfont=sf]{subfig}
\usepackage{textcomp}
\usepackage{stfloats}
\usepackage{tabularx}
\usepackage{xcolor}
\usepackage{url}
\usepackage{verbatim}
\usepackage{multirow}
\usepackage{acronym}
\usepackage{subcaption}
\usepackage{graphicx}	
\usepackage{stfloats}
\usepackage{soul}
\usepackage{cancel}
\usepackage{cite}
\def\BibTeX{{\rm B\kern-.05em{\sc i\kern-.025em b}\kern-.08em
T\kern-.1667em\lower.7ex\hbox{E}\kern-.125emX}}
\usepackage{balance}
\usepackage{multirow}
\usepackage{hhline}
\input{AcronymsListFinal}
\acresetall

\begin{document}
\title{Mission Capability-Aware UAV Swarm Coordination for Autonomous Disaster Response}
%%%%%%%%%%%%%%%%%%%%%%%%%%%
\author{Mudassar Shakeel, Taha Mahmood, Ahmed~Naeem, Muhammad Hamza Jehangir, El Mehdi Amhoud~\IEEEmembership{Senior Member,~IEEE}, Syed Ali Hassan~\IEEEmembership{Senior Member,~IEEE}\\This work has been submitted to the IEEE for possible publication. Copyright may be transferred without notice, after which this version may no longer be accessible.
\thanks{Mudassar Shakeel, Taha Mahmood, Muhammad Hamza Jehangir, Ahmed~Naeem, and Syed Ali Hassan are with the School of Electrical Engineering and Computer Science, National University of Sciences and Technology, 44000, Islamabad, Pakistan. (email: \{mshakeel.bee23seecs, tmahmood.bee23seecs, ahmed.naeem,
mjehangir.bee23seecs, ali.hassan\}@seecs.edu.pk).\\ 
El Mehdi Amhoud is with the College of Computing, University Mohammed VI Polytechnic, Benguerir 43150, Morocco (email: elmehdi.amhoud@um6p.ma).\\
}
}
\markboth{Journal of \LaTeX\ Class Files,~Vol.~14, No.~8, February~2025}%
{Shell \MakeLowercase{\textit{et al.}}: Bare Demo of IEEEtran.cls for IEEE Journals}
%%%%%%%%%%%%%%%%%%%%%%%%%%%
\maketitle 
\begin{abstract}
This paper proposes an \ac{AI}-driven mission capability-aware \ac{UAV} swarm coordination framework for autonomous post-disaster search and rescue. An integrated sensing and communication-enabled master high-altitude platform station senses the disaster environment, detects victims and hazards, and constructs adaptive rescue regions based on spatially varying mission demands. A \ac{MCR} quantifies regional capability demand, while a normalized \ac{MCS} characterizes the real-time capability of individual \acp{UAV}. A capability-aware assignment mechanism matches UAV capabilities to regional requirements while minimizing deployment distance. An \ac{MCS}-based reallocation mechanism dynamically transfers UAVs from capability-surplus to capability-deficient regions while preserving donor sufficiency. A hierarchical deep
reinforcement learning framework with proximal policy optimization performs high-level region selection, while \ac{MCR}--\ac{MCS} matching enables feasible UAV-level reallocation. Finally, joint trajectory optimization considers travel time, distance, propulsion energy, operational capability, obstacle avoidance, and inter-\ac{UAV} collision avoidance. Theoretical analysis and simulations demonstrate improved mission capability satisfaction, resource efficiency, and disaster-response performance.
\end{abstract}
\acresetall
%%%%%%%%%%%%%%%%%%%%%%%%%%%%%%%%%%%%%%%%%%%%%%%%%%%%%%%%%%%%%%%
\section{Introduction}
\subsection{Background}
Natural disasters such as earthquakes, floods, hurricanes, and wildfires frequently destroy terrestrial communication infrastructure, severely limiting situational awareness, wireless connectivity, and coordination during emergency \ac{SAR} operations~\cite{1,2}. Rapid disaster assessment and efficient deployment of rescue resources are therefore essential for improving victim survival rates~\cite{3}. Due to their rapid deployment, mobility, and aerial sensing capabilities, \ac{UAV} swarms are promising for post-disaster response~\cite{4}. Furthermore, the evolution of \ac{6G} wireless networks and \ac{ISAC} enables \acp{UAV} to simultaneously perform environmental sensing and wireless communication using shared spectrum and hardware resources~\cite{5}. This capability provides real-time situational awareness while maintaining wireless connectivity, making \ac{ISAC}-enabled \ac{UAV} swarms an attractive platform for autonomous disaster management~\cite{6}.

Despite these advantages, effective coordination of \ac{UAV} swarms in highly dynamic disaster environments remains challenging. Disaster conditions are spatially heterogeneous, and victim density, environmental conditions, disaster intensity, communication quality, and accessibility may vary significantly across different locations~\cite{6.5}. Consequently, assigning \acp{UAV} using fixed partitioning or a single criterion, such as travel distance or residual battery energy, may create an imbalance between rescue requirements and assigned \acp{UAV}. Highly affected regions may receive insufficient resources, while less critical regions may be over-served, leading to inefficient utilization of limited \ac{UAV} resources~\cite{6.6}.

Another important challenge arises from the heterogeneous and time-varying operational capabilities of individual \acp{UAV}. During a rescue mission, the operational capability of each \ac{UAV} continuously changes due to battery depletion, communication degradation, sensing performance, workload, payload status, and remaining flight time~\cite{13,15}. Therefore, an initially feasible \ac{UAV} assignment may become inadequate as mission conditions evolve. Moreover, \ac{AI}-based disaster perception may produce uncertain predictions under challenging sensing conditions. Directly using unreliable perception results for disaster partitioning and resource allocation can consequently degrade subsequent mission-planning decisions. These challenges motivate the development of an autonomous \ac{UAV} swarm coordination framework that jointly considers adaptive disaster partitioning, \ac{AI} prediction reliability, heterogeneous \ac{UAV} capabilities, regional rescue requirements, and continuously changing mission conditions.
%%%%%%%%%%%%%%%%%%%%%%%%%%%%%%%%%%%%%%%%%%%%%%%%%%%%%%%%%%%%%%%
\subsection{Related Works}
Existing studies have explored \ac{UAV}-assisted disaster response for
sensing, victim detection, emergency communication, and information
collection, as well as swarm coordination using optimization, heuristic, and multi-agent task allocation~\cite{8}. However, most resource allocation strategies rely on travel distance, residual battery energy, communication quality, or mission priorities~\cite{9}, without explicitly quantifying regional rescue capability requirements. Consequently, a single criterion may assign \acp{UAV} with insufficient capability to demanding disaster regions.

Disaster-area partitioning and \ac{UAV} path-planning techniques have been investigated to improve coverage and timely information collection. Work has explored spatial partitioning, critical-node-aware coverage, and trajectory optimization for \ac{UAV} swarm operations in disaster environments~\cite{7}. However, conventional uniform grid-based partitioning may not accurately represent the heterogeneous nature of real disaster environments, where victim density, environmental conditions, disaster intensity, and accessibility vary significantly across different locations~\cite{11,12}. Although adaptive partitioning better represents spatial variations, existing approaches generally do not explicitly associate each disaster region with a quantitative.

\Ac{AI} and \ac{DRL} have increasingly been applied to \ac{UAV} task allocation, trajectory planning, resource management, and autonomous decision making ~\cite{14,16}. These approaches can improve adaptability by learning from dynamically changing environmental conditions. However, many existing \ac{AI}-assisted disaster-response frameworks primarily optimize individual objectives such as energy consumption, flight distance, coverage, or task completion time. They generally do not explicitly model the capability required to accomplish different rescue missions ~\cite{17}. Furthermore, \Ac{AI} perception uncertainty is often neglected before using perception results for resource allocation, potentially leading to unreliable mission-planning decisions under challenging sensing conditions.

Another important research direction is the use of \ac{ISAC}-enabled \ac{UAV} systems for simultaneous environmental sensing and wireless communication~\cite{18}. \Ac{ISAC} provides an attractive foundation for disaster-response applications because \acp{UAV} can continuously acquire environmental information while maintaining wireless connectivity. Existing \ac{ISAC}-enabled \ac{UAV} studies have mainly focused on sensing performance, communication reliability, beamforming, trajectory optimization, and joint resource management. However, the integration of \ac{ISAC}-based perception with confidence-aware disaster partitioning and mission capability-aware \ac{UAV} coordination remains insufficiently explored~\cite{19}.

Furthermore, most existing \ac{UAV} allocation approaches assume relatively stable or homogeneous \ac{UAV} capabilities. In practical
disaster-response missions, however, the operational capability of each
\ac{UAV} varies over time due to battery consumption, sensing and communication conditions, workload, payload status, and remaining flight
time~\cite{13,15}. Thus, an initially feasible assignment may become unsuitable as mission conditions evolve, highlighting the need to jointly consider evolving regional requirements and time-varying \ac{UAV} capabilities. Therefore, existing studies typically address individual aspects of disaster response, such as perception, communication, area partitioning,
trajectory planning, or \ac{UAV} task allocation. However, the joint integration of reliable \ac{AI} perception, adaptive rescue-region formation, mission capability modeling, and capability-aware \ac{UAV} coordination remains insufficiently explored. In particular, continuously matching evolving regional rescue requirements with the time-varying capabilities of individual \acp{UAV} remains an open challenge.

To address these challenges, this paper proposes an \ac{AI}-driven mission capability-aware \ac{UAV} swarm coordination framework for autonomous disaster response. The framework employs an \ac{ISAC}-enabled master \ac{HAPS} and multiple rescue \acp{UAV}. The master \ac{HAPS} performs global sensing, wireless communication, \ac{AI}-based disaster perception, adaptive disaster-region formation, and mission coordination. The framework combines confidence-aware perception, adaptive disaster partitioning, \ac{MCS}, capability-aware \ac{UAV} assignment, dynamic capability reallocation, and \ac{DRL}-based mission coordination to adapt \ac{UAV} deployment to evolving disaster conditions.
%%%%%%%%%%%%%%%%%%%%%%%%%%%%%%%%%%%%%%%%%%%%%%%%%%%%%%%%%%%%%%%
\subsection{Contributions}
Compared with conventional \ac{UAV} coordination approaches that primarily rely on travel distance, residual battery energy, or heuristic task allocation, the proposed framework introduces mission capability as a central decision metric for autonomous disaster-response coordination. The main contributions of this paper are summarized as follows:
\begin{itemize}
\item We propose an \ac{ISAC}-enabled autonomous \ac{UAV} swarm architecture
with a master \ac{HAPS} that jointly performs global sensing, communication,
and mission~coordination.
\item We develop a confidence-aware \ac{AI} perception and adaptive disaster
partitioning mechanism that estimates spatially varying victim density,
environmental conditions, disaster intensity, rescue demand, and perception
confidence, and forms heterogeneous rescue regions based on validated mission requirements.
\item We introduce the \ac{MCR} and effective \ac{MCS} framework, together with a capability-aware assignment and dynamic reallocation mechanism that matches \ac{UAV} capabilities to regional requirements while considering traveling
cost and preserving donor-region sufficiency.
\item We integrate a \ac{DRL}-based high-level coordination strategy with
the \ac{MCR}--\ac{MCS} framework, where \ac{PPO} selects capability-deficient regions. The \ac{MCR}--\ac{MCS} matching determines feasible \ac{UAV} reallocation, and trajectory optimization generates physically feasible paths while considering travel time, distance, energy consumption, obstacle avoidance, and inter-\ac{UAV}
collision avoidance.
\end{itemize}
%%%%%%%%%%%%%%%%%%%%%%%%%%%%%%%%%%%%%%%%%%%%%%%%%%%%%%%%%%%
\begin{figure*}
\centering 
\resizebox{1.65\columnwidth}{!}{
\includegraphics{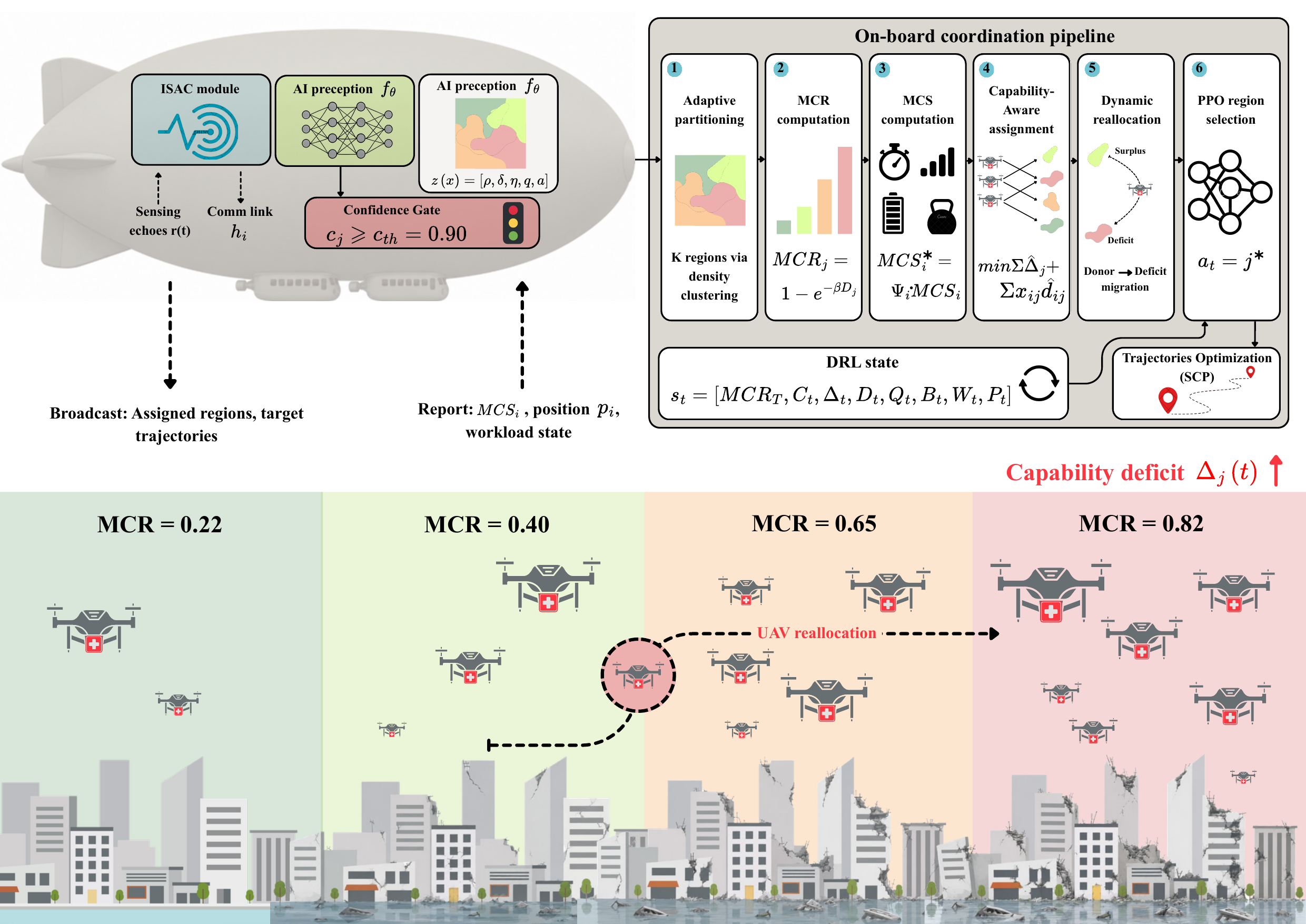}}
\caption{System architecture of the proposed mission capability-aware UAV swarm coordination framework.}
\label{overall_architecture}
\end{figure*}
%%%%%%%%%%%%%%%%%%%%%%%%%%%%%%%%%%%%%%%%%%%%%%%%%%%%%%%%%%%%%%
\section{System Model} \label{SYS2}
Consider a post-disaster \ac{SAR} scenario where terrestrial communication infrastructure is severely damaged. For disaster sensing and rescue coordination, an autonomous aerial network consisting of a rescue \ac{UAV} and a master \ac{HAPS} with \ac{ISAC} capability is used. The master \ac{HAPS} processes \ac{ISAC}
observations to partition the mission area, estimate regional \ac{MCR} and \ac{UAV} \ac{MCS}, and identify capability deficits as illustrated in Fig.~\ref{overall_architecture}. A \ac{PPO}-based module
then selects target regions for capability-aware \ac{UAV} reallocation, followed by trajectory optimization for the assigned rescue \acp{UAV}.
%%%%%%%%%%%%%%%%%%%%%%%%%
\subsection{Network Architecture}
\par The post-disaster area is modeled as a bounded 2D region $\Omega \subset \mathbb{R}^2$. The rescue \ac{UAV} swarm is denoted by $\mathcal{U}=\{U_1,U_2,\dots,U_N\}$, where $N$ is the total number of rescue \acp{UAV}. A single master \ac{HAPS}, with higher computational
capability, longer endurance, and wide-area coverage of $\Omega$, oversees the mission by performing (i) \ac{ISAC}-based sensing, (ii) \ac{AI}-based disaster analysis, (iii) mission planning, and (iv) rescue \ac{UAV} swarm coordination.

\par The 3D position of the master \ac{HAPS} at time $t$ is denoted by $p_0(t) = [x_0(t), y_0(t), h_0]^T$, where $x_0(t)$ and $y_0(t)$ denote its coordinates and $h_0$ is the flying altitude of the master \ac{HAPS}. The position of the $i$-th rescue \ac{UAV} is denoted by $p_i(t) = [x_i(t), y_i(t), h_i]^T, \quad i = 1, \dots, N,$ where $x_i(t)$ and $y_i(t)$ represent the coordinates and $h_i$ is the altitude of the $i$-th rescue \ac{UAV}. To reflect realistic \ac{UAV} mobility, the velocity of every rescue \ac{UAV} $v_i(t)$, is constrained by $\|v_i(t)\| \leq V_{\max}$, where $V_{\max}$ is the maximum admissible \ac{UAV} speed. The master \ac{HAPS} and the rescue \ac{UAV} swarm exchange sensing information, mission commands, and status updates through a dedicated wireless link that is jointly established with the \ac{ISAC} module described in Section II-B. This link is crucial for the capability-aware coordination described later in this work because it enables the master \ac{HAPS} to communicate mission-planning choices to the swarm and to continuously receive updated \ac{UAV} status.
%%%%%%%%%%%%%%%%%%%%%%%%%
\subsection{\ac{ISAC}-Based Communication and Sensing Model}
\par To avoid separate sensing and communication subsystems, the master \ac{HAPS} employs a unified \ac{ISAC} waveform $s(t)$ for victim
detection, obstacle identification, environmental monitoring, and communication with the rescue \ac{UAV} swarm. The \ac{HAPS} illuminates the
disaster area using $s(t)$ and processes reflected echoes from $M$ targets, including victims, obstacles, and damaged structures. The received multi-target sensing signal is modeled as~\cite{20}
%%%%%%%%%%%%%%%%
\begin{equation}
r(t) =
\sum_{m=1}^{M}
a_m s(t-d_m)
e^{j2\pi f_{D,m}t}+n(t),
\label{eq:sensing_signal}
\end{equation}
%%%%%%%%%%%%%%%%
where $a_m$ denotes the complex reflection coefficient of the $m$-th target, $d_m$ denotes the propagation delay, and
$f_{D,m}$ denotes the Doppler frequency due to the relative
motion between the target and the master \ac{HAPS}, and $n(t)$ represents additive receiver noise. Estimating $d_m$ provides target-range information, while $f_{D,m}$ provides target-motion information. These estimates are used by the master \ac{HAPS} to localize detected objects and update the global disaster map introduced
in Section~II-C.

For communication, the master \ac{HAPS} establishes wireless links with the rescue \acp{UAV}. The time-varying channel between the master \ac{HAPS} and the $i$-th rescue \ac{UAV} is modeled as $h_i(t)=\sqrt{b_i(t)}g_i(t),$ where $g_i(t)$ represents the small-scale fading coefficient and $b_i(t)$ represents the large-scale channel power gain, modeled as  $b_i(t)=b_od_i(t)^{-a_{\mathrm{pl}}}$ where $d_i(t)$ is the distance between the master \ac{HAPS} and the $i$-th rescue \ac{UAV}, $a_\mathrm{pl}$ is the path-loss exponent, and $b_o$ is the reference channel power gain. Accordingly, the achievable communication rate is
$R_i(t)=B\log_2\left(1+\frac{P_t|h_i(t)|^2}{\sigma_n^2}\right),$ where $B$ denotes the communication bandwidth, $P_t$ is the master \ac{HAPS} transmit power, and $\sigma_n^2$ is the total receiver noise power over $B$.

To capture spatially varying communication conditions in the disaster map and region formation, communication quality is evaluated independently of rescue \ac{UAV} assignment. For an arbitrary spatial
location $\mathbf{x}$ within the disaster area, the \ac{HAPS}-to-location channel is modeled as $H(\mathbf{x})=\sqrt{b(\mathbf{x})}g(\mathbf{x}),$
where $b(\mathbf{x})$ denotes the large-scale channel power gain and $g(\mathbf{x})$ represents the corresponding small-scale fading coefficient. The achievable communication rate at location
$\mathbf{x}$ is given by $R(\mathbf{x}) =B\log_2\left(1+\frac{P_t|H(\mathbf{x})|^2}{N_0}\right).$ The normalized spatial communication quality is defined as
%%%%%%%%%%%%%%%%
\begin{equation}
q(\mathbf{x})=
\min\left(
\frac{R(\mathbf{x})}{R_{\max}},\,1
\right),
\label{eq:spatial_communication_quality}
\end{equation}
%%%%%%%%%%%%%%%%
where $R_{\max}$ denotes the reference communication rate used for normalization. Consequently, $0\leq q(\mathbf{x})\leq1$, where
$q(\mathbf{x})=0$ represents extremely poor communication conditions, while $q(\mathbf{x})=1$ indicates that the achievable rate meets or exceeds the reference rate $R_{\max}$. Let $\mathcal{X}_j$ denote the set of spatial sampling points within disaster region $R_j$. Assuming equal importance for all sampling points, the regional communication quality is obtained as $q_j =\frac{1}{|\mathcal{X}_j|}\sum_{\mathbf{x}\in\mathcal{X}_j}q(\mathbf{x}), \mathcal{X}_j\neq\emptyset.$ Thus, $q_j\in[0,1]$ measures the spatial communication quality of region
$R_j$ and is used for adaptive region formation and rescue-demand estimation independently of \ac{UAV} assignment. After assignment, the instantaneous communication quality of the $i$-th rescue \ac{UAV} is defined as $Q_i(t)=\min\left(\frac{R_i(t)}{R_{\max}},\,1\right).$ The $Q_i(t)$ is incorporated into the \ac{MCS} and communication-feasibility constraint, whereas $q(\mathbf{x})$ and $q_j$ characterize the spatial
communication environment before \ac{UAV} assignment.
%%%%%%%%%%%%%%%%%%%%%%%%%
\subsection{Global Disaster Map Construction and Confidence-Aware \ac{AI} Disaster Assessment}
\par The sensing outputs collected through the \ac{ISAC} module at the master \ac{HAPS} are aggregated into a global disaster map $\mathcal{M}=\{x,z(x)\}$, $x\in\Omega$, where each spatial location $x$ is associated with the disaster feature vector
$z(x)=[p(x),s(x),\eta(x),q(x),a(x)]^T,$ where $p(x)$ denotes victim density, $s(x)$ disaster intensity, $\eta(x)$ environmental difficulty (e.g., debris, fire, or flooding severity), $q(x)$ local communication quality derived from \eqref{eq:spatial_communication_quality}, and $a(x)$ physical accessibility. The map $\mathcal{M}$ provides a unified spatial representation of the disaster environment and serves as input to \ac{AI} perception and mission-planning stages. The raw sensing observations are noisy and incomplete, particularly during early rescue operations. The master \ac{HAPS} feeds $Y(x)$ into a trained \ac{AI} model $f_{\theta}(\cdot)$, which outputs an estimated feature vector as
%%%%%%%%%%%%%%%%
\begin{equation}
\hat{\mathbf{z}}(x)
=f_{\theta}\big(Y(x)\big)
=\left[
\hat{p}(x),
\hat{s}(x),
\hat{\eta}(x),
\hat{q}(x),
\hat{a}(x)
\right]^{T}.
\label{eq:mud}
\end{equation}
%%%%%%%%%%%%%%%%
In addition to $\hat{\mathbf{z}}(x)$, the \ac{AI} model produces a pointwise confidence score $c(x)\in[0,1]$, which quantifies the reliability of the estimate at location $x$. For each adaptive disaster region $R_j$, the regional confidence is summarized by the spatially averaged confidence, $c_j=\frac{1}{|R_j|}\int_{R_j}c(x)\,dx$. The adaptive disaster partitioning is considered reliable for rescue-resource allocation only when $c_j \geq c_{\mathrm{th}},~ c_{\mathrm{th}}=0.90 ~\forall j,$ where $c_{\mathrm{th}}$ denotes a predefined confidence threshold. If $c_j<c_{\mathrm{th}}$ for any region, the master \ac{HAPS} continues acquiring \ac{ISAC} observations and updates \ac{AI} perception results before committing the rescue \ac{UAV} swarm to corresponding mission plan.

\par The evolution of the \ac{AI} confidence during successive \ac{ISAC} sensing
iterations is illustrated in Fig.~\ref{fig:confidence_adaptive_partitioning}(a).
During initial sensing iterations, the confidence score is relatively low because the master \ac{HAPS} has limited environmental observations and incomplete information about the disaster state. As additional \ac{ISAC}
observations are acquired, the confidence score increases and eventually reaches the predefined reliability level of $0.90$. Once confidence criterion is satisfied, the corresponding \ac{AI} perception results can be used for adaptive disaster partitioning and subsequent \ac{UAV} resource allocation. This confidence-gating mechanism prevents premature mission decisions based on unreliable environmental information and improves robustness of the proposed disaster-response framework.
%%%%%%%%%%%%%%%%%%%%%%%%%
\subsection{Adaptive Disaster Partitioning and Capability Modeling} \label{2D}
\par Once the confidence condition is satisfied, the master \ac{HAPS} partitions the validated disaster map into a set of adaptive rescue regions denoted as
%%%%%%%%%%%%%%%%
\begin{equation}
\mathcal{R} = \{R_1,R_2,\ldots,R_K\},
\label{eq:r}
\end{equation}
where $K$ denotes the number of regions obtained through a density-based clustering method.
%%%%%%%%%%%%%%%%%%%%%%%%%%%%%%%%
\begin{figure*}
\centering 
\subfloat[]{
  \includegraphics[width=42.5mm,height=40mm]{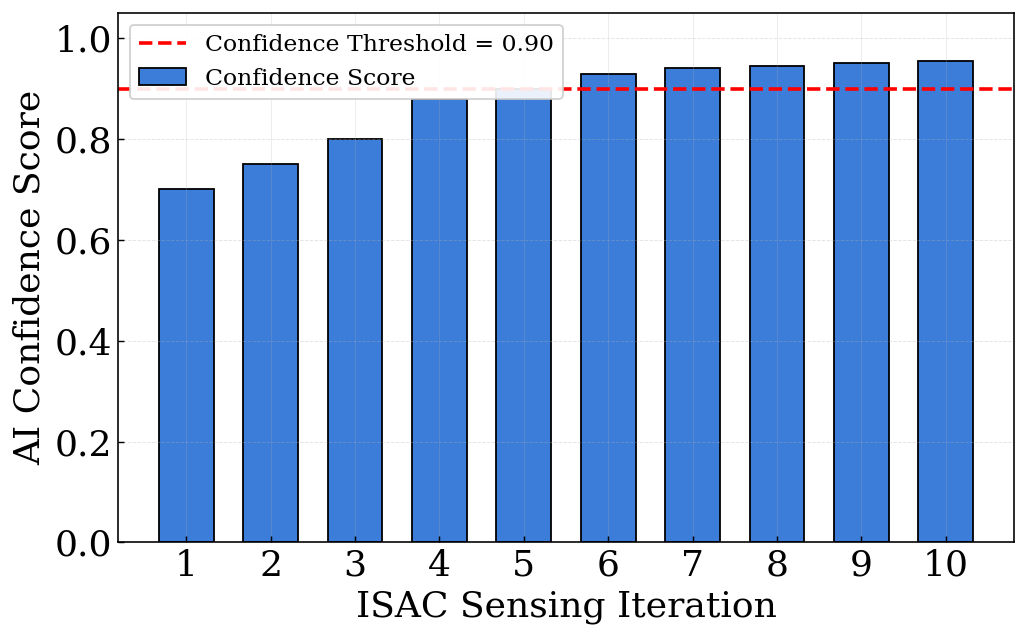}
}
\subfloat[]{
  \includegraphics[width=42.5mm,height=40mm]{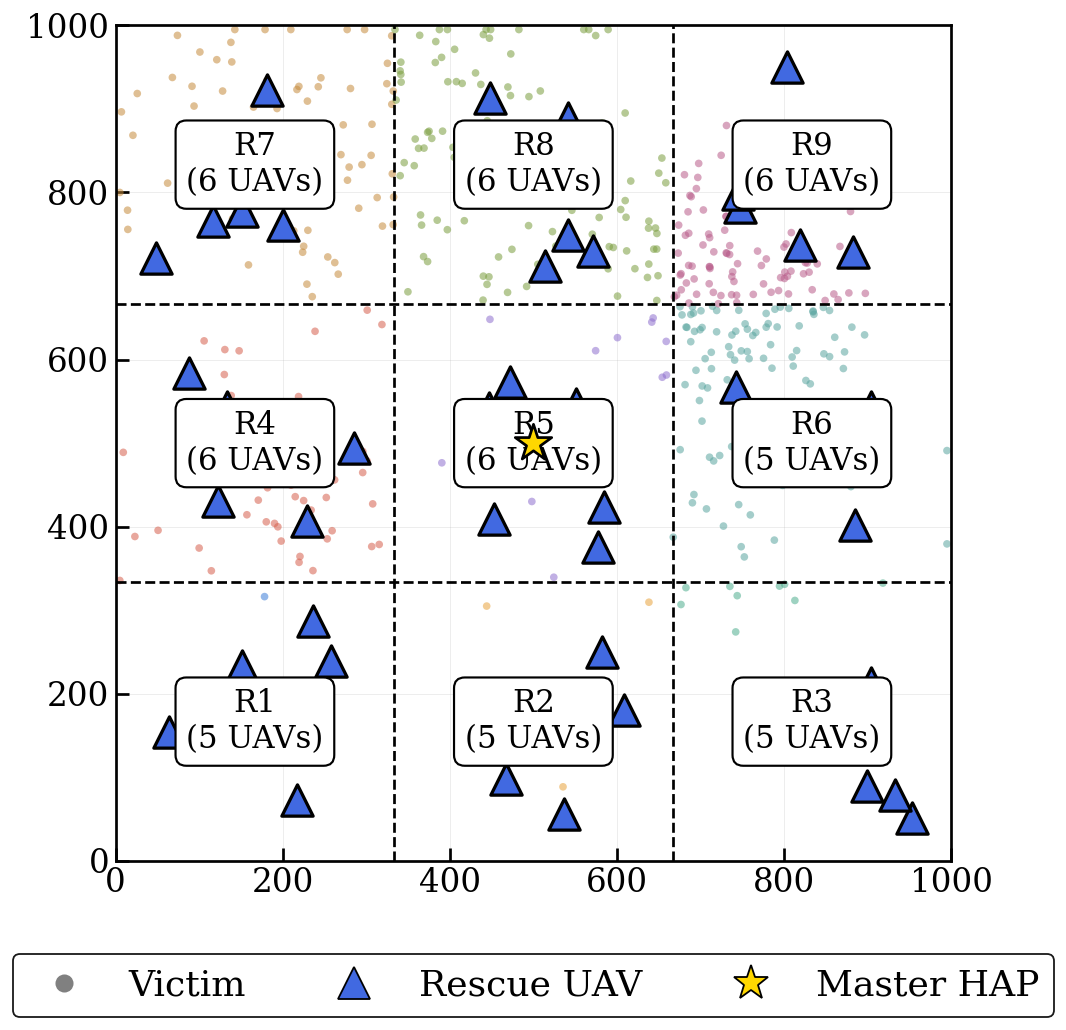}
}
\subfloat[]{
  \includegraphics[width=42.5mm,height=40mm]{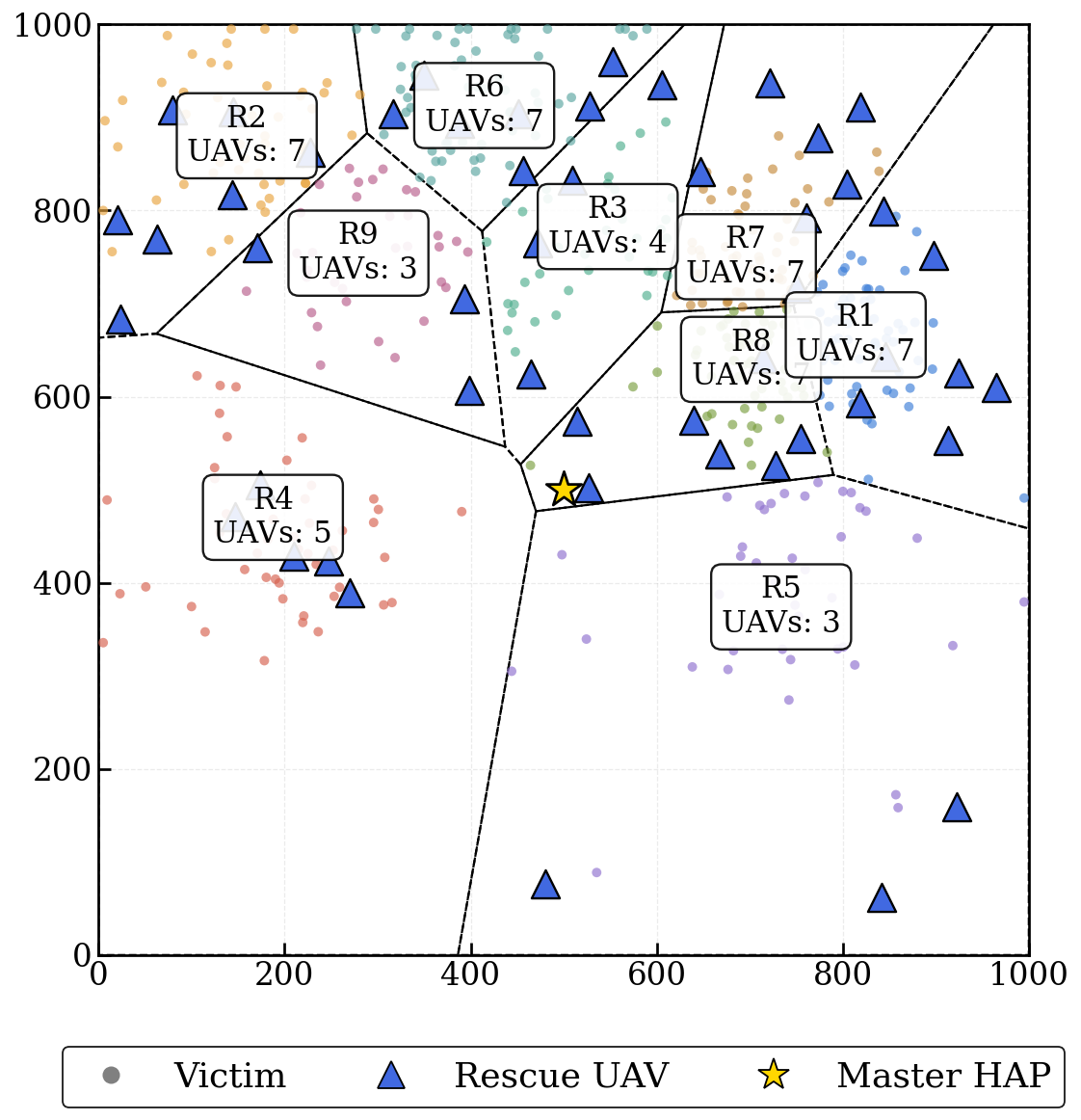}
}
\subfloat[]{
  \includegraphics[width=42.5mm,height=40mm]{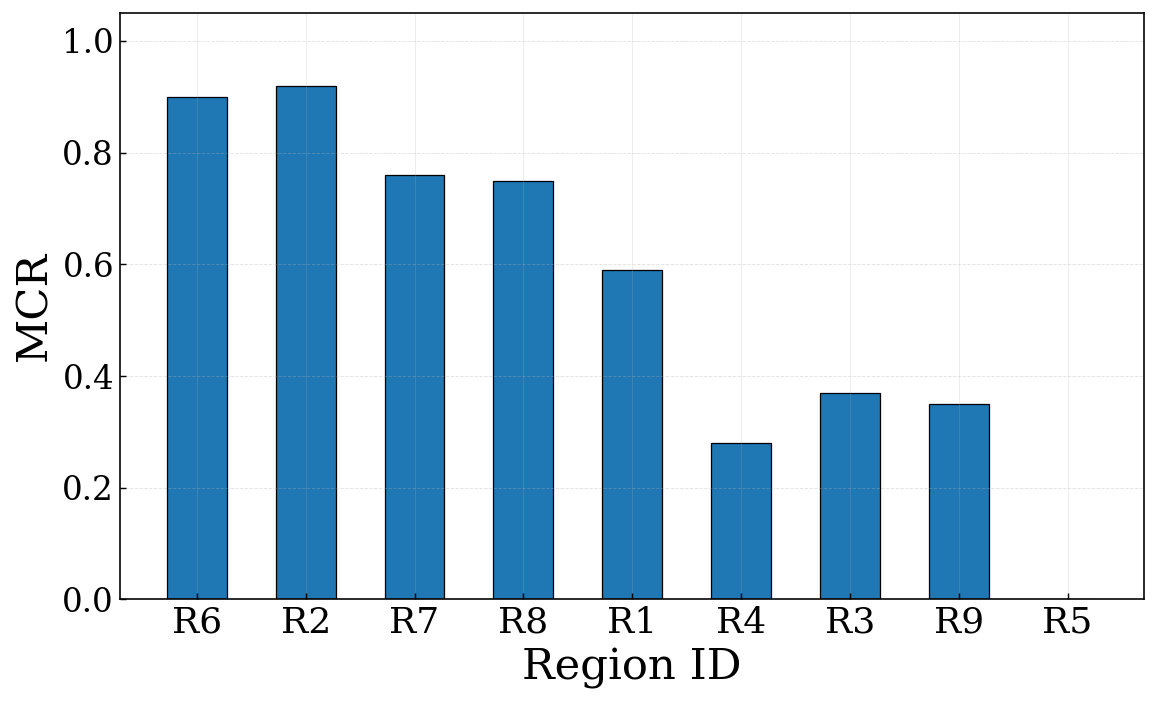}
}
\caption{(a) Conventional uniform-grid partitioning, (b) proposed partitioning with \ac{MCR}-based \ac{UAV} allocation, (c) \ac{AI} perception confidence across \ac{ISAC} iterations, and (d) initial \ac{MCR} across adaptive rescue regions before \ac{DRL}-based \ac{UAV} reallocation.}
\label{fig:confidence_adaptive_partitioning}
\end{figure*}
%%%%%%%%%%%%%%%%%%%%%%%%%%%%%%%%
The difference between conventional uniform partitioning~\cite{11,12} and the proposed adaptive partitioning is illustrated in Fig.~\ref{fig:confidence_adaptive_partitioning}(b) and
Fig.~\ref{fig:confidence_adaptive_partitioning}(c), respectively. Fig.~\ref{fig:confidence_adaptive_partitioning}(b) represents the conventional equal-grid partitioning, in which the disaster area is divided into spatially uniform regions. In contrast, Fig.~\ref{fig:confidence_adaptive_partitioning}(c) shows the proposed adaptive partitioning, where rescue regions are formed according to spatially varying disaster features and rescue demand. Consequently, different regions can obtain different \ac{MCR}, enabling the subsequent \ac{UAV} allocation to account for heterogeneous mission requirements. Unlike a fixed grid, the proposed approach groups spatially neighboring locations with similar estimated disaster features, resulting in regions that better represent heterogeneous disaster conditions.

\par For each $R_j$, the normalized regional feature values are obtained by averaging corresponding location-level features.
%%%%%%%%%%%%%%%%
\begin{equation}
\hat{p}_j =
\frac{1}{|R_j|}\sum_{u\in R_j}\hat{p}_u,\qquad
\hat{s}_j =
\frac{1}{|R_j|}\sum_{u\in R_j}\hat{s}_u,
\end{equation}
%%%%%%%%%%%%%%%%
\begin{equation}
\hat{\eta}_j =
\frac{1}{|R_j|}\sum_{u\in R_j}\hat{\eta}_u,~
\hat{q}_j =
\frac{1}{|R_j|}\sum_{u\in R_j}\hat{q}_u,~
\hat{a}_j =
\frac{1}{|R_j|}\sum_{u\in R_j}\hat{a}_u.
\end{equation}
%%%%%%%%%%%%%%%%
The rescue demand of region $R_j$ is defined as
%%%%%%%%%%%%%%%%
\begin{equation}
D_j =
w_p\hat{p}_j +
w_s\hat{s}_j +
w_{\eta}\hat{\eta}_j +
w_q(1-\hat{q}_j) +
w_a(1-\hat{a}_j),
\label{eq:region_demand}
\end{equation}
%%%%%%%%%%%%%%%%
where $w_p,w_s,w_{\eta},w_q,w_a\geq0$ are the corresponding
weights satisfying $w_p+w_s+w_{\eta}+w_q+w_a=1$.
Thus, higher disaster severity and lower communication quality or accessibility increase rescue demand. The demand is mapped to the bounded \ac{MCR} as ${MCR}_j=1-\exp(-k_r D_j),$ where $k_r>0$ controls the mapping sensitivity. Each $i$-th rescue \ac{UAV} maintains the normalized operational state vector $\mathbf{s}_i=
[\tilde{B}_i,\tilde{S}_i,\tilde{Q}_i,\tilde{F}_i,\tilde{W}_i]^T,$ where all state variables lie in $[0,1]$. The \ac{MCS} of the $i$-th \ac{UAV} is defined in \eqref{eq:mcs}. Higher battery, sensing,
communication, and endurance levels increase mission capability, whereas higher workload reduces it through $(1-\tilde{W}_i)$.
%%%%%%%%%%%%%%%%%%%%%%%%%%%%%%%%%%%%%%%%%%%%%%%%%%%%%%%%%%%
\section{Problem Formulation} \label{problem}
Following adaptive disaster partitioning, the master \ac{HAPS} autonomously allocates rescue \acp{UAV} according to regional \ac{MCR}. It continuously evaluates regional demand and \ac{UAV} operational capability to formulate capability-aware assignment and dynamic capability reallocation.
%%%%%%%%%%%%%%%%%%%%%%%%%%%%%%%%%%%%%%%%%%%%%%%%%%%%%%%%%%%
\subsection{\ac{AI}-Based Rescue Demand and \ac{MCR}}
\label{subsec:MCR}
For each spatial location $\mathbf{x}\in\Omega$, the \ac{ISAC}-enabled master \ac{HAPS} obtains the sensing observation $\mathbf{Y}(\mathbf{x})$. The \ac{AI} model estimates the disaster-related features and prediction confidence $c(\mathbf{x})\in[0,1]$ as in \eqref{eq:mud}. Before adaptive region formation, the estimated features are normalized using min--max normalization as $\tilde{x}(\mathbf{x})=\frac{\hat{x}(\mathbf{x})-x_{\min}}
{x_{\max}-x_{\min}},$ yielding the normalized feature vector $\tilde{\mathbf{z}}(\mathbf{x})
=\left[\tilde{p}(\mathbf{x}),\tilde{s}(\mathbf{x}),
\tilde{\eta}(\mathbf{x}),\tilde{q}(\mathbf{x}),
\tilde{a}(\mathbf{x})\right]^T.$
For each adaptive rescue region $R_j$, the regional confidence is computed by averaging the pointwise confidence values as $c_j=\frac{1}{|R_j|}
\int_{R_j} c(\mathbf{x})\,d\mathbf{x}.$ The confidence validation is then performed at the regional level according to
%%%%%%%%%%%%%%%%%%%%%%%
\begin{equation}
\mathcal{P}_j=
\begin{cases}
1, & c_j\geq c_{\mathrm{th}},\\
0, & c_j<c_{\mathrm{th}},
\end{cases}
\qquad c_{\mathrm{th}}=0.90.
\end{equation}
%%%%%%%%%%%%%%%%%%%%%%%
Only regions satisfying $\mathcal{P}_j=1$ are used for subsequent mission planning, while low-confidence regions are re-observed in the next sensing cycle. The validated observations are partitioned into
$\mathcal{R}=\{R_1,R_2,\ldots,R_K\}$, and the normalized features within each region are averaged to obtain the regional feature vector $\mathbf{z}_j =
\left[\hat{p}_j,
\hat{s}_j,\hat{\eta}_j,\hat{q}_j,\hat{a}_j\right]^T,$
where the components denote the normalized victim density, disaster intensity, environmental difficulty, communication quality, and accessibility of region $R_j$, respectively. The regional rescue-demand score $D_j$ is defined in \eqref{eq:region_demand}, where severe disaster conditions and poor communication or accessibility increase the rescue demand. The corresponding  is defined as
%%%%%%%%%%%%%%%%%%%%%%%
\begin{equation}
\mathrm{MCR}_j
= 1-e^{-k_r D_j},
\label{eq:mcr}
\end{equation}
%%%%%%%%%%%%%%%%%%%%%%%
where $k_r>0$ controls the sensitivity of the capability requirement to regional rescue demand. Since $D_j\geq0$, the resulting satisfies $0\leq\mathrm{MCR}_j<1$, with larger values indicating greater required operational capability. As illustrated in
Fig.~\ref{fig:confidence_adaptive_partitioning}(d), the adaptive rescue regions exhibit heterogeneous capability requirements, with $R_2$ and $R_6$ showing the highest \ac{MCR} values. This variation reflects the
spatially nonuniform rescue demand captured by the proposed formulation and enables capability-aware prioritization of rescue \acp{UAV}.
%%%%%%%%%%%%%%%%%%%%%%%%%%%%%%%%%%%%%%%%%%%%%%%%
\subsection{Mission Capability Score}
\label{subsec:mcs}
All state variables are normalized to $[0,1]$.
To avoid double counting the battery contribution, the intrinsic capability score is first calculated from the non-battery operational factors as
%%%%%%%%%%%%%%%%
\begin{equation}
\mathrm{MCS}_i =
a_S\tilde{S}_i +
a_Q\tilde{Q}_i +
a_F\tilde{F}_i +
a_W(1-\tilde{W}_i).
\label{eq:mcs}
\end{equation}
%%%%%%%%%%%%%%%%
where $a_S,a_Q,a_F,a_W\geq0$ and
$a_S+a_Q+a_F+a_W=1$. Since all state variables are normalized to $[0,1]$, the intrinsic capability score satisfies $0\leq \mathrm{MCS}_i\leq1$.
The remaining battery level is incorporated through a nonlinear battery capability factor, defined as $\Psi_i =
1-\exp(-k_g\tilde{B}_i),$ where $k_g>0$ controls the sensitivity of the capability factor to the remaining battery level. This function satisfies $0\leq\Psi_i<1$ and increases monotonically with $\tilde{B}_i$.
The effective \ac{MCS}, denoted by $\mathrm{MCS}_i^*$, is therefore defined as
%%%%%%%%%%%%%%%%
\begin{equation}
\mathrm{MCS}_i^{*}
=
\Psi_i \mathrm{MCS}_i,
\qquad 0\leq \mathrm{MCS}_i^{*}<1.
\label{eq:effective_mcs}
\end{equation}
%%%%%%%%%%%%%%%%
The effective \ac{MCS} represents each \ac{UAV}'s capability relative to a common reference $C_{\mathrm{ref}}>0$. Although individual scores are bounded by unity, the aggregate capability may exceed unity under divisible rescue workloads and approximately additive \ac{UAV} contributions.

\par The $\mathrm{MCS}_i^*$ is periodically updated according to the latest \ac{UAV} state information and transmitted to the master \ac{HAPS}. This enables the master \ac{HAPS} to maintain an up-to-date capability representation of the rescue \ac{UAV} swarm for capability-aware assignment and dynamic mission reconfiguration. The compatibility between the $i$-th rescue \ac{UAV} and rescue region $R_j$ is measured using the matching utility
%%%%%%%%%%%%%%%%
\begin{equation}
U_{ij}=
\exp\left(
-\left|\mathrm{MCS}_i^*-\mathrm{MCR}_j\right|
-\lambda_d\tilde{d}_{ij}
\right),
\label{eq:mcs_matching_utility}
\end{equation}
%%%%%%%%%%%%%%%%
where $\tilde{d}_{ij}$ is the normalized traveling distance
between the $i$-th rescue \ac{UAV} and region $R_j$,
and $\lambda_d\geq0$ controls the importance of traveling
distance. A larger $U_{ij}$ indicates better capability
compatibility and shorter travel distance between the
\ac{UAV} and the corresponding rescue region.
%%%%%%%%%%%%%%%%%%%%%%%%%%%%%%%%%%%%%%%%%%%%%%%%%%%%%
\subsection{Capability-Aware \ac{UAV} Assignment}
\label{subsec:initial_assignment}
\par After computing the \ac{MCR} of each
adaptive rescue region and the \ac{MCS} of
each rescue \ac{UAV}, the master \ac{HAPS} determines the initial \ac{UAV} assignment. Unlike conventional uniform-grid partitioning and distance-based \ac{UAV} assignment strategies as illustrated in Fig.~\ref{fig:confidence_adaptive_partitioning}(b), the proposed approach uses the adaptive regional structure shown in Fig.~\ref{fig:confidence_adaptive_partitioning}(c) for capability-aware \ac{UAV} assignment. The proposed assignment considers the compatibility between the capability requirement of each region and the operational capability of the available \acp{UAV}. Let the binary assignment variable be defined as
%%%%%%%%%%%%%%%%
\begin{equation}
x_{ij} =
\begin{cases}
1, & \text{if UAV } U_i \text{ is assigned to region } R_j,\\
0, & \text{otherwise},
\end{cases}
\label{eq:initial_assignment}
\end{equation}
%%%%%%%%%%%%%%%%
where $i=1,\ldots,N$ and $j=1,\ldots,K$.
The total rescue capability allocated to region $R_j$ is obtained by summing the effective \ac{MCS} of its assigned \acp{UAV} on a common reference scale as
%%%%%%%%%%%%%%%%
\begin{equation}
C_j =
\sum_{i=1}^{N} x_{ij}\mathrm{MCS}_i^*.
\label{eq:initial_allocated_capability}
\end{equation}
%%%%%%%%%%%%%%%%
The aggregate capability $C_j$ may exceed unity and satisfies the regional requirement when $C_j\geq\mathrm{MCR}_j$. When enough UAV capacity is available, the initial assignment must meet the adaptive rescue regions' capability requirements. UAV exclusivity, regional UAV coverage, capability sufficiency, and binary assignment constraints represent some of the assignment requirements. The resulting assignment provides the initial deployment of the rescue \acp{UAV}. During mission execution, changes in regional demand and \ac{UAV} operational capability are handled separately by the dynamic mission capability reallocation mechanism.
%%%%%%%%%%%%%%%%%%%%%%%%%%%%%%%%%%%%%%%%%%%%%%%%%%%%%%%%%%%
\subsection{Multi-Objective Mission Optimization}
The objective of the proposed framework is to determine a \ac{UAV} assignment and trajectory planning strategy that maximizes rescue effectiveness while minimizing mission completion time, energy consumption, rescue delay, communication degradation, and workload imbalance. Let $\mathbf{X}=[x_{ij}]$ denote the \ac{UAV} assignment matrix, where $x_{ij}=1$ if rescue \ac{UAV} $U_i$ is assigned to adaptive region $R_j$, and $x_{ij}=0$ otherwise. The trajectory set of the rescue \acp{UAV} is represented by $\Gamma=\left\{\mathbf{p}_1(t),
\mathbf{p}_2(t),\ldots,\mathbf{p}_N(t)\right\}.$ The proposed framework simultaneously considers seven performance objectives as $\mathbf{F}(\mathbf{X},\Gamma)
=\left[\hat{T},\hat{E},\hat{D},\widehat{Co},-\widehat{P}_s,\widehat{CD},\widehat{WI}
\right]^T,$ where $T$, $D$, $\widehat{Cov}$, $\widehat{P}_s$, $\widehat{CD}$, and $\widehat{WI}$ denotes the mission completion time, total energy consumption of the rescue \ac{UAV}, victim coverage, communication degradation, and workload imbalance (the hat notation indicates normalized objective values).

\par The resulting multi-objective optimization problem is formulated as $\min_{\mathbf{X},\Gamma}\mathbf{F}(\mathbf{X},\Gamma)$. Since these objectives are generally conflicting, a feasible solution $(\mathbf{X}^{*},\Gamma^{*})$ is Pareto optimal if no other
feasible solution can improve one objective without degrading at least one of the remaining objectives. For practical implementation, the multi-objective formulation is converted into a weighted scalar optimization problem $\min_{\mathbf{X},\Gamma}
\sum_{m=1}^{7}\lambda_m F_m(\mathbf{X},\Gamma),$ where $\sum_{m=1}^{7}\lambda_m=1, \lambda_m\geq0.$ The resulting scalar optimization problem is therefore as
%%%%%%%%%%%%%%%%
\begin{equation}
\begin{aligned}
\min_{\mathbf{X},\Gamma,\boldsymbol{\xi}}\quad
&\lambda_I\hat{T}
+\lambda_2\hat{E}
+\lambda_3\hat{D}
-\lambda_4\widehat{Cov}\\
&-\lambda_5\widehat{P}_s
+\lambda_6\widehat{CD}
+\lambda_7\widehat{WI}
+\lambda_8\sum_{j=1}^{K}\xi_j,
\end{aligned}
\end{equation}
\begin{align}
\mathrm{s.t.} &
\sum_{i=1}^{N}
x_{ij}\mathrm{MCS}_i^{*}
+\xi_j
\geq
\mathrm{MCR}_j,
&& \forall j,
\label{eq:capability_constraint}
\\
&
\xi_j\geq0,
&& \forall j,
\label{eq:unmet_capability_constraint}
\\
&
\sum_{j=1}^{K}
x_{ij}
\leq 1,
&& \forall i,
\label{eq:assignment_constraint}
\\
&
x_{ij}\in\{0,1\},
&& \forall i,j,
\label{eq:binary_constraint}
\\
&
\left\|
\dot{\mathbf{p}}_i(t)
\right\|_2
\leq V_{\max},
&& \forall i,t,
\label{eq:velocity_constraint}
\\
&
\left\|
\mathbf{p}_i(t)-\mathbf{o}_m
\right\|_2
\geq r_m^{\mathrm{safe}},
&& \forall i,m,t,
\label{eq:obstacle_constraint}
\\
&
B_i(t)\geq B_{\min},
&& \forall i,t,
\label{eq:battery_constraint}
\\
&
Q_i(t)\geq Q_{\min},
&& \forall i,t,
\label{eq:communication_constraint}
\\
&
F_i(t)\geq F_{\min},
&& \forall i,t,
\label{eq:endurance_constraint}
\end{align}
%%%%%%%%%%%%%%%%
where the constraint in \eqref{eq:capability_constraint}
ensures that the total operational capability assigned to each adaptive rescue region, together with any unmet capability $\xi_j$, satisfies its \ac{MCR}. The coefficient $\lambda_8>0$ penalizes unmet regional capability. The constraint in
\eqref{eq:assignment_constraint} ensures that each rescue \ac{UAV} is assigned to at most one adaptive region during a scheduling interval. The binary constraint in \eqref{eq:binary_constraint} defines the
\ac{UAV}-to-region assignment decisions. The velocity constraint in \eqref{eq:velocity_constraint} limits \ac{UAV} motion to the allowable maximum speed, while the obstacle-avoidance constraint in \eqref{eq:obstacle_constraint} prevents \ac{UAV} trajectories from entering unsafe obstacle regions. The battery constraint in \eqref{eq:battery_constraint} guarantees that each \ac{UAV} maintains a minimum residual energy level throughout the mission. The constraint in
\eqref{eq:communication_constraint} maintains the required
communication quality between the master \ac{HAPS} and each rescue \ac{UAV}.
Finally, the flight-endurance constraint in
\eqref{eq:endurance_constraint} ensures that the remaining flight time of each \ac{UAV} remains above the minimum operational threshold.
%%%%%%%%%%%%%%%%%%%%%%%%%%%%%%%%%%%%%%%%%%%%%%%%%%%%%%%%%%%%%%%%%%%%%%%%%%
\section{Proposed Method} \label{method}
%%%%%%%%%%%%%%%%%%%%%%%%%%%%%%%%%%%%%%%%%%%%%%%%%%%%%%%%%%%%%%%%%%%%%
Building on the \ac{MCR} of each adaptive rescue region and the \ac{MCS} of each rescue \ac{UAV}, we propose an \ac{AI}-driven mission capability-aware \ac{UAV} swarm coordination framework executed by the \ac{ISAC}-enabled master \ac{HAPS}. It comprises four stages: (i) capability-aware initial \ac{UAV} assignment, (ii) dynamic mission capability reallocation, (iii) \ac{DRL}-based adaptive coordination, and (iv) joint trajectory optimization. At each scheduling interval, the master \ac{HAPS} updates regional \ac{MCR} from confidence-validated \ac{AI} perception and receives the current \ac{MCS} of each \ac{UAV}. It then performs capability-aware assignment and reallocation, while the \ac{DRL} coordinator adapts high-level decisions to changing mission conditions and the trajectory optimizer generates feasible, energy-efficient \ac{UAV} trajectories.
%%%%%%%%%%%%%%%%%%%%%%%%%%%%%%%%%%%%%%%%%%%%%%%%%%%%%%%%%%%%%%%%%%%%%
\subsection{Capability-Aware Initial \ac{UAV} Assignment}
\label{subsec:capability_aware_assignment}

\par Following the \ac{UAV} assignment formulation, the master \ac{HAPS} initially deploys the rescue \ac{UAV} fleet by matching the capability requirements of the adaptive rescue regions with the effective capabilities of available \acp{UAV}. Unlike nearest-\ac{UAV} assignment,
the proposed approach prioritizes compatibility between the
\ac{MCR} of each region and the effectiveness of \ac{MCS} for each \ac{UAV}, while considering traveling distance as a secondary criterion. The assignment at mission time $t$ is represented by $\mathbf{X}(t)=[x_{ij}(t)]\in\{0,1\}^{N\times K},$
%%%%%%%%%%%%%%%%
where $x_{ij}(t)=1$ indicates that \ac{UAV} $U_i$ is assigned to region
$R_j$. The capability deficit of region $R_j$ is defined as
%%%%%%%%%%%%%%%%
\begin{equation}
\Delta_j(t)=
\left[
\mathrm{MCR}_j(t)-
\sum_{i=1}^{N}x_{ij}(t)\mathrm{MCS}_i^{*}(t)
\right]^+,
\label{eq:capability_deficit}
\end{equation}
%%%%%%%%%%%%%%%%
where $[z]^+=\max(z,0)$. Thus, $\Delta_j(t)=0$ indicates sufficient allocated capability, while $\Delta_j(t)>0$ denotes an unmet requirement. Let $\widehat{\Delta}_j(t)$ and $\widehat{d}_{ij}(t)$ denote the normalized capability deficit and the distance between \ac{UAV} $U_i$ and the centroid of region $R_j$, respectively. The capability-aware assignment is
%%%%%%%%%%%%%%%%
\begin{equation}
\min_{\mathbf{X}(t)}
\;
\alpha\sum_{j=1}^{K}\widehat{\Delta}_j(t)
+
\beta\sum_{i=1}^{N}\sum_{j=1}^{K}
x_{ij}(t)\widehat{d}_{ij}(t),
\label{eq:capability_aware_objective}
\end{equation}
%%%%%%%%%%%%%%%%
where $\alpha+\beta=1$ and $\alpha,\beta\geq0$ determine the relative
importance of capability sufficiency and travel distance. The first term promotes capability-feasible assignments, while the second reduces unnecessary \ac{UAV} travel. Thus, a higher-capability \ac{UAV} may be preferred over a closer \ac{UAV} when it better reduces the regional deficit. The resulting $\mathbf{X}(t)$ defines the initial \ac{UAV} distribution. If fleet capability is insufficient, the soft-deficit formulation minimizes unmet requirements. As regional demand and \ac{UAV} capability evolve, the master \ac{HAPS} updates these states and triggers dynamic capability reallocation when required.
%%%%%%%%%%%%%%%%%%%%%%%%%%%%%%%%%%%%
\subsection{Dynamic Mission Capability Reallocation}
The disaster evolves during \ac{SAR}, regional rescue
demand and \ac{UAV} operational capability vary with newly detected victims, battery consumption, communication degradation, and environmental conditions. Thus, an initially feasible assignment may become inadequate. The master \ac{HAPS} therefore periodically updates regional and \ac{UAV} \ac{MCS} using the latest \ac{ISAC} observations. The capability gap of $R_j$ is $G_j(t)=\mathrm{MCR}_j(t)-C_j(t),$ where $G_j(t)>0$ indicates that the allocated rescue capability is insufficient, $G_j(t)=0$ indicates capability balance, and $G_j(t)<0$ indicates that the region has surplus capability. For capability-deficit minimization, the non-negative deficit is defined as $\Delta_j(t)=\left[G_j(t)\right]^+=\left[\mathrm{MCR}_j(t)-C_j(t)\right]^+.$ Accordingly, capability status of $R_j$ is
%%%%%%%%%%%%%%%%
\begin{equation}
\Phi_j(t)
=
\begin{cases}
\mathrm{Deficit}, & G_j(t)>0,\\
\mathrm{Balanced}, & G_j(t)=0,\\
\mathrm{Surplus}, & G_j(t)<0.
\end{cases}
\label{eq:capability_status}
\end{equation}
%%%%%%%%%%%%%%%%
When $\Delta_j(t)>0$, $R_j$ requires additional rescue capability. Conversely, $R_k$ can act as a donor when
$C_k(t)>MCR_k(t).$ Let the set of \acp{UAV} currently assigned to donor region $R_k$ be defined as
$\mathcal{U}_k(t)=\left\{
i:x_{ik}(t)=1\right\}.$ A \ac{UAV} $U_i\in\mathcal{U}_k(t)$ can be transferred from the surplus region $R_k$ to the deficit $R_j$ only if its removal does not violate the capability requirement of the donor region. Therefore, the feasible donor set is
%%%%%%%%%%%%%%%%
\begin{equation}
\mathcal{F}_{jk}(t)
=\left\{
i\in\mathcal{U}_k(t):
C_k(t)-MCS_i^{*}(t)\geq MCR_k(t)
\right\}.
\label{eq:feasible_donor_set}
\end{equation}
%%%%%%%%%%%%%%%%
For each feasible \ac{UAV}, the distance to the centroid of the deficit region is calculated as $D_{ij}(t)=
\left\|\mathbf{p}_i(t)-\mathbf{c}_j(t)
\right\|_2,$ where $\mathbf{p}_i(t)$ denotes the position of \ac{UAV} $U_i$ and $\mathbf{c}_j(t)$ denotes the centroid of region $R_j$. To compare candidates \acp{UAV} under different mission conditions, the distance is normalized as
%%%%%%%%%%%%%%%%
\begin{equation}
\widetilde{D}_{ij}(t)
=\frac{D_{ij}(t)}
{\displaystyle
\max_{\ell\in\mathcal{F}_{jk}(t)}
D_{\ell j}(t)+\epsilon},
\label{eq:normalized_reallocation_distance}
\end{equation}
%%%%%%%%%%%%%%%%
where $\epsilon>0$ is a small numerical constant.
The master \ac{HAPS} selects the \ac{UAV} for migration by minimizing a capability-, distance-, and workload-aware selection cost
%%%%%%%%%%%%%%%%
\begin{equation}
i^{\star}
=\arg\max_{i\in\mathcal{F}_{jk}(t)}
\left[b_c \mathrm{MCS}_i^*(t)-
b_d \widetilde{D}_{ij}(t)
-b_w \widetilde{W}_i(t)\right].
\label{eq:selected_uav}
\end{equation}
%%%%%%%%%%%%%%%%
where $\mathcal{F}_{jk^{*}}(t)$ denotes the feasible donor \ac{UAV} set, $\widetilde{D}_{ij^{*}}(t)$ is the normalized distance between \ac{UAV} $U_i$ and the target region, $\widetilde{W}_i(t)$ denotes the normalized workload of \ac{UAV} $U_i$, while $b_c$, $b_d$, and $b_w$ are nonnegative selection weights satisfying
$b_c+b_d+b_w=1.$ The selected \ac{UAV} is removed from the donor region and assigned to the deficit region as
$x_{i^{*}k}(t+1)=0, x_{i^{*}j}(t+1)=1.$ Consequently, the allocated capabilities of both regions are updated as
%%%%%%%%%%%%%%%%
\begin{align}
C_k(t+1)
&=
C_k(t)-MCS_{i^{*}}^{*}(t),
\label{eq:donor_capability_update}
\\
C_j(t+1)
&=
C_j(t)+MCS_{i^{*}}^{*}(t).
\label{eq:receiver_capability_update}
\end{align}
%%%%%%%%%%%%%%%%
The reallocation procedure is repeated for the remaining deficit regions until all feasible capability deficits are sufficiently reduced or no feasible donor \ac{UAV} remains. Through this dynamic reallocation mechanism, the master \ac{HAPS} continuously adapts the \ac{UAV}
distribution to evolving regional rescue requirements while
preserving the capability sufficiency of donor regions and improving mission-level resource utilization.
%%%%%%%%%%%%%%%%%%%%%%%%%%%%%%%%%%%%%%%%%%%%%%%%%%%%%%%%%%%%%%%%%%%%%
\subsection{Joint Trajectory Optimization}
%%%%%%%%%%%%%%%%%%%%%%%%%%%%%%%%%%%%%%%%%%%%%%%%%%%%%%%%%%%%%%%%%%%%%

Following capability-aware assignment and dynamic capability reallocation, the master \ac{HAPS} determines feasible trajectories for the assigned rescue \acp{UAV}. The trajectory optimizer considers rescue
urgency, traveling distance, operational capability, energy
consumption, communication connectivity, obstacle avoidance, and inter-\ac{UAV} collision avoidance.
Let the trajectory of rescue \ac{UAV} $U_i$ assigned to region $R_j$ be represented by
$\Gamma_i =\left\{\mathbf{p}_i(0),
\mathbf{p}_i(1),\ldots,\mathbf{p}_i(L)\right\},$
where $L$ denotes the trajectory planning horizon and
$\mathbf{p}_i(l)\in\mathbb{R}^{3}$ represents the 3D
position of \ac{UAV} $U_i$ at way point $l$.
The corresponding discrete-time velocity is given by
%%%%%%%%%%%%%%%%%%%%
\begin{equation}
\mathbf{v}_i(l) = \frac{\mathbf{p}_i(l+1)-\mathbf{p}_i(l) }{ \Delta t }, \qquad l=0,\ldots,L 1,
\label{eq:trajectory_velocity}
\end{equation}
%%%%%%%%%%%%%%%%%%%%
where $\Delta t$ denotes the trajectory sampling interval.
The total traveling distance of \ac{UAV} $U_i$ is calculated as $D_i=\sum_{l=0}^{L-1}\left\|
\mathbf{p}_i(l+1)-\mathbf{p}_i(l)\right\|_2,$ and the corresponding traveling time is $T_i=L\Delta t.$ For computational simplicity, the propulsion energy consumed by
\ac{UAV} $U_i$ is approximated by
%%%%%%%%%%%%%%%%%%%%
\begin{equation}
E_i=\sum_{l=0}^{L-1}
\left(
c_1\|\mathbf{v}_i(l)\|_2^3+c_2
\right)\Delta t,
\label{eq:propulsion_energy}
\end{equation}
%%%%%%%%%%%%%%%%%%%%
where $c_1$ and $c_2$ are platform-dependent propulsion
coefficients. To jointly account for traveling time, distance, energy consumption, and operational capability, the trajectory optimization for \ac{UAV} $U_i$ is formulated as
%%%%%%%%%%%%%%%%%%%%
\begin{equation}
\begin{aligned}
\min_{\Gamma_i}
\quad
J_i^{\mathrm{traj}}
=
&
\lambda_T \frac{T_i}{T_{\max}}
+
\lambda_D \frac{D_i}{D_{\max}}
+
\lambda_E \frac{E_i}{E_{\max}}
\\
&
+
\lambda_C
\left(
1-\widetilde{\mathrm{MCS}}_i(t)
\right),
\end{aligned}
\label{eq:trajectory_optimization}
\end{equation}
%%%%%%%%%%%%%%%%%%%%
where $\lambda_T$, $\lambda_D$, $\lambda_E$, and $\lambda_C$ are nonnegative weighting coefficients satisfying
$\lambda_T+\lambda_D+\lambda_E+\lambda_C=1$. The trajectory optimization follows the mission-level feasibility constraints, including velocity, obstacle avoidance, inter-\ac{UAV} collision avoidance, and battery limits. Here, $T_i$, $D_i$, and $E_i$ denote trajectory duration, traveled distance, and propulsion energy consumption, respectively, while $\widetilde{\mathrm{MCS}}_i(t)$ denotes the normalized operational capability of \ac{UAV} $U_i$. The \ac{MCS} term preserves \ac{UAV} capability, whereas the time, distance, and energy terms promote rapid and resource-efficient movement.
Due to the nonlinear obstacle- and collision-avoidance constraints, the trajectory problem is a nonlinear constrained optimization problem. Sequential convex programming iteratively linearize these constraints around the current trajectory and solves the resulting convex subproblem until convergence.
%%%%%%%%%%%%%%%%%%%%%%%%%%%%%%%%%%%%%%%%%%%%%%%%%%%%%%%%%%%%%%%%%%%%%
\subsection{\ac{DRL}-Based Adaptive Mission Coordination}
Although capability-aware assignment and dynamic \ac{MCR}--\ac{MCS} reallocation are effective at individual scheduling intervals, evolving disaster conditions can make previous allocations inadequate. Changes in victim distribution, environmental conditions, \ac{UAV} battery, communication quality, and regional demand therefore require long-term adaptive coordination. Accordingly, a \ac{DRL} agent at the master \ac{HAPS} performs high-level mission coordination by identifying regions requiring additional rescue capability, while the \ac{MCR}--\ac{MCS} mechanism selects feasible \acp{UAV} for reallocation. The resulting decisions are then converted into physically feasible \ac{UAV} movements through trajectory control.
The dynamic coordination problem is formulated as Markov Decision Process $
\mathcal{M}=(\mathcal{S},\mathcal{A},\mathcal{P},\mathcal{R},\gamma),$
where $\mathcal{S}$ denotes the state space, $\mathcal{A}$ denotes the action space, $\mathcal{P}$ represents the state-transition probability, $\mathcal{R}$ denotes the reward function, and $\gamma\in[0,1)$ is the discount factor.
%%%%%%%%%%%%%%%%%%%%%%%%%%%%%%%%%%%%%%%%%%%%%%%%%%%%%%%%%%%%%%%%%%%%%%%%%%
\subsubsection{State Space}
At scheduling interval $t$, the master \ac{HAPS} constructs a global state
representation using the latest \ac{ISAC} observations, regional mission
requirements, and \ac{UAV} operational conditions. The state is defined as $\mathbf{s}_{t}=\left[\mathbf{MCR}_{t},\mathbf{C}_{t},\boldsymbol{\Delta}_{t},\mathbf{D}_{t},\mathbf{Q}_{t},\mathbf{B}_{t},\mathbf{W}_{t},\mathbf{P}_{t}\right]^{T},$
where $\mathbf{MCR}_{t}$ contains the  of all adaptive rescue regions, $\mathbf{C}_{t}$ denotes the rescue capability currently allocated to each region, and $\boldsymbol{\Delta}_{t}$ represents the corresponding regional capability deficits. Moreover, $\mathbf{D}_{t}$ denotes the regional rescue
demand, while $\mathbf{Q}_{t}$ represents communication quality. Furthermore, $\mathbf{B}_{t}$ and $\mathbf{W}_{t}$ denote the residual battery and
workload states of the rescue \acp{UAV}, respectively, and $\mathbf{P}_{t}$
contains their current positions.
For adaptive region $R_j$, the total rescue capability currently
available at time $t$ is defined as
$ C_j(t) = \sum_{i=1}^{N} x_{ij}(t)MCS_i^{*}(t).$
The capability deficit of $R_j$ is consequently expressed as
$\Delta_j(t) = \max\left( MCR_j(t)-C_j(t),0 \right).$
The total capability deficit of the disaster area is given by
$\Delta_{\mathrm{tot}}(t) = \sum_{j=1}^{K}\Delta_j(t).$
This quantity provides a direct measure of the mismatch between the
rescue capability required by the adaptive regions and the capability
currently available from the assigned \acp{UAV}.

%%%%%%%%%%%%%%%%%%%%%%%%%%%%%%%%%%%%%%%%%%%%%%%%%%%%%%%%%%%%%%%%%%%%%%%%%%
\subsubsection{High-Level Action Space}

Unlike direct \ac{UAV}-control formulations, where the \ac{DRL} agent selects
individual \ac{UAV} trajectories, the proposed framework adopts a hierarchical
action structure. The \ac{PPO} agent makes a high-level region-selection
decision,
%%%%%%%%
$a_t=j^{*},  j^{*}\in\{1,\ldots,K\},$
where $j^{*}$ denotes the adaptive disaster region requiring additional
rescue capability. Thus, the \ac{PPO} agent selects the region requiring
additional capability rather than directly selecting a rescue \ac{UAV} and
its trajectory, reducing the action-space dimension and focusing the
learned policy on long-term mission-level coordination. After selecting
$R_{j^{*}}$, the capability-aware \ac{MCR}--\ac{MCS} reallocation mechanism
identifies a rescue \ac{UAV} from a feasible donor region $R_k$, where
%%%%%%%%%%%%%%%%
\begin{equation}
C_k(t)-MCS_i^{*}(t)
\geq MCR_k(t),
\label{eq:drl_donor_constraint}
\end{equation}
%%%%%%%%%%%%%%%%
for candidate \ac{UAV} $U_i$. For each feasible candidate, the migration cost
is evaluated using the capability-, distance-, and workload-aware
selection criterion in \eqref{eq:selected_uav}. Hence, the hierarchical
mechanism separates long-term region selection from \ac{UAV}-level capability
matching: the \ac{PPO} agent selects $R_{j^{*}}$, while the
\ac{MCR}--\ac{MCS} mechanism selects a feasible $U_i$ while preserving
the donor region's capability requirement.

%%%%%%%%%%%%%%%%%%%%%%%%%%%%%%%%%%%%%%%%%%%%%%%%%%%%%%%%%%%%%%%%%%%%%%%%%%
\subsubsection{State Transition and Dynamic Reallocation}

Once \ac{UAV} $U_{i^{*}}$ is selected, it is removed from its donor region
$R_k$ and assigned to the \ac{PPO}-selected target region $R_{j^{*}}$.
Accordingly,
$x_{i^{*}k}(t+1)=0,$ and $x_{i^{*}j^{*}}(t+1)=1.
$
The regional capabilities are consequently updated as
$C_k(t+1) = C_k(t)-MCS_{i^{*}}^{*}(t),$ and 
$ C_{j^{*}}(t+1) = C_{j^{*}}(t)+MCS_{i^{*}}^{*}(t).
$
Following the reallocation, the master \ac{HAPS} receives updated \ac{ISAC}
observations and recomputes the  of each region and the \ac{MCS} of every
rescue \ac{UAV}. Consequently, the state evolves according to
$ \mathbf{s}_{t+1} = f\left( \mathbf{s}_{t},a_t,\mathbf{e}_{t} \right), $
where $\mathbf{e}_{t}$ represents the dynamically changing disaster
environment. The state transition therefore captures variations in
victim demand, environmental conditions, \ac{UAV} battery, communication
quality, workload, and regional capability requirements.

\begin{figure}[t]
\subfloat[]{
  \includegraphics[width=42mm,height=40mm]{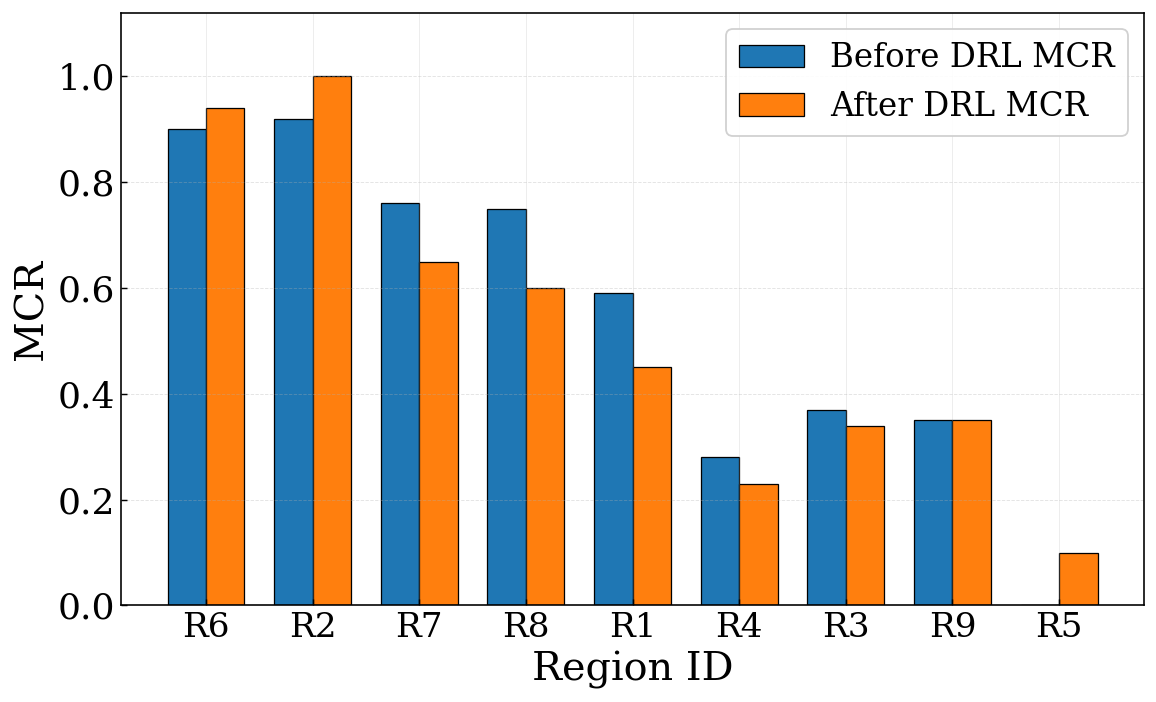}
}
\subfloat[]
{
  \includegraphics[width=42mm,height=40mm]{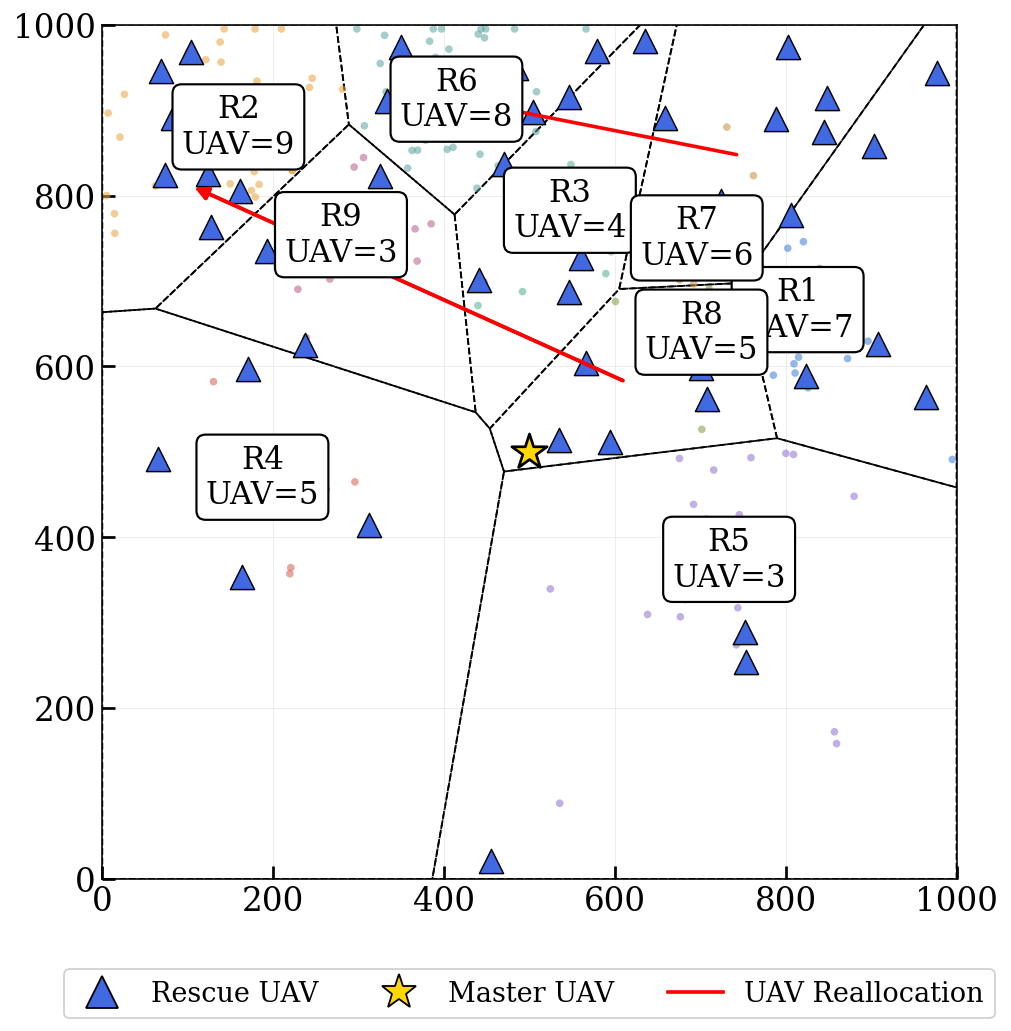}
}
\caption{MCR and UAV allocation before and after DRL-based
coordination: (a) initial and updated regional MCR values, and (b) UAV assignment after dynamic reallocation.}
\label{fig:drl_mcr_reallocation}
\end{figure}

%%%%%%%%%%%%%%%%%%%%%%%%%%%%%%%%%%%%%%%%%%%%%%%%%%%%%%%%%%%%%%%%%%%%%%%%%%
\textbf{ Update and DRL-Based Reallocation:}
Fig.~\ref{fig:drl_mcr_reallocation} illustrates the interaction between
updated regional mission requirements and \ac{DRL}-based \ac{UAV} reallocation.
Fig.~\ref{fig:drl_mcr_reallocation}(a) compares the initial and updated values, where the updates are generated by new \ac{ISAC} observations
and \ac{AI}-based disaster perception, including changes in victim
distribution, environmental conditions, and rescue demand, rather than
by the \ac{DRL} agent.
Fig.~\ref{fig:confidence_adaptive_partitioning}(c) shows the \ac{UAV} distribution under the updated mission state, where changed regional requirements may
create capability deficits. The \ac{PPO} agent observes the updated
regional capability states and selects the region requiring additional
rescue capability. The \ac{MCR}--\ac{MCS} mechanism then identifies a feasible
\ac{UAV} from a donor region while ensuring that the donor continues to
satisfy its own capability requirement. The resulting distribution is
shown in Fig.~\ref{fig:drl_mcr_reallocation}(b). Thus, the proposed
hierarchical framework uses updated \ac{ISAC}/\ac{AI} information to determine, \ac{PPO} to select the target region, and the \ac{MCR}--\ac{MCS} mechanism to determine the feasible \ac{UAV} reallocation.

%%%%%%%%%%%%%%%%%%%%%%%%%%%%%%%%%%%%%%%%%%%%%%%%%%%%%%%%%%%%%%%%%%%%%%%%%%
\subsubsection{Reward Function}

The reward function encourages the \ac{PPO} policy to reduce regional
capability deficits while minimizing rescue delay and resource
consumption. Since the objectives have different physical units, their
normalized forms are used. The immediate reward is defined as
%%%%%%%%%%%%%%%%%%%%%%%%
\begin{equation}
\begin{aligned}
r_t
={}&
-\lambda_1\overline{\Delta}_{\mathrm{tot}}(t)
-\lambda_2\overline{E}(t)
-\lambda_3\overline{D}(t)
+\lambda_4\overline{\mathrm{Cov}}(t)\\
&-\lambda_5\overline{CD}(t)
-\lambda_6\overline{C}_{\mathrm{realloc}}(t),
\end{aligned}
\label{eq:drl_reward}
\end{equation}
%%%%%%%%%%%%%%%%%%%%%%%%
where $\overline{\Delta}_{\mathrm{tot}}(t)$, $\overline{E}(t)$,
$\overline{D}(t)$, $\overline{\mathrm{Cov}}(t)$, $\overline{CD}(t)$, and
$\overline{C}_{\mathrm{realloc}}(t)$ denote the normalized total capability deficit, energy consumption, rescue delay, victim coverage,
communication degradation, and \ac{UAV} reallocation cost, respectively.
The coefficients $\lambda_1,\ldots,\lambda_6\geq0$ control their relative importance. Higher capability deficit, energy consumption, rescue delay, communication degradation, and reallocation cost reduce the reward, whereas increased victim coverage improves it.

%%%%%%%%%%%%%%%%%%%%%%%%%%%%%%%%%%%%%%%%%%%%%%%%%%%%%%%%%%%%%%%%%%%%%%%%%%
\subsubsection{\ac{PPO}-Based Policy Optimization}

The \ac{PPO} agent seeks a policy $\pi_{\theta}$ that maximizes the expected
discounted cumulative reward over a scheduling horizon $H$
%%%%%%%%%%%%%%%%%%%%%%%%%%%%%%%%
\begin{equation}
J(\theta)
= \mathbb{E}_{\pi_{\theta}} \left[ \sum_{t=0}^{H-1}
g_d^{t}r_t \right],
\end{equation}
%%%%%%%%%%%%%%%%%%%%%%%%%%%%%%%%
where $\theta$ denotes the parameters of the policy network.
The \ac{PPO} policy is optimized using the clipped surrogate objective,
$ L^{\mathrm{CLIP}}(\theta) = \mathbb{E}_{t} [ \min
( p_t(\theta)\hat{A}_t, \operatorname{clip} \left( p_t(\theta), 1-\epsilon, 1+\epsilon \right) \hat{A}_t )], $
where
$ p_t(\theta) = \frac{ \pi_{\theta}(a_t|\mathbf{s}_t) }{
\pi_{\theta_{\mathrm{old}}}(a_t|\mathbf{s}_t) },$
$\hat{A}_t$ is the estimated advantage function and $\epsilon$ is the \ac{PPO} clipping coefficient. The policy parameters are updated as
$ \theta_{n+1} = \theta_n + \alpha_{\theta}
\nabla_{\theta} L^{\mathrm{CLIP}}(\theta_n),$
where $\alpha_{\theta}$ is the policy learning rate.
At each scheduling interval, the master \ac{HAPS} updates the \ac{ISAC} observations, regional, \ac{UAV} \ac{MCS}, capability
deficits, and state $\mathbf{s}_t$. The trained \ac{PPO} policy selects the region requiring additional capability, while the
\ac{MCR}--\ac{MCS} mechanism selects a feasible donor \ac{UAV} subject
to (\ref{eq:drl_donor_constraint}). The resulting reallocation is executed through the \ac{UAV} mobility-control layer, separating mission-level region selection from \ac{UAV}-level capability-aware reallocation and enabling continuous adaptation to evolving disaster conditions.

%%%%%%%%%%%%%%%%%%%%%%%%%%%%%%%%%%%%%%%%%%%%%%%%%%%%%%%%%%%%%%%%%%%%%%%%%%
\section{Performance Analysis} \label{perf}

This section theoretically analyzes the proposed framework from five perspectives: mission capability satisfaction, rescue delay, energy consumption, system stability, and computational complexity. The analysis incorporates AI-based disaster perception, dynamic MCR estimation, capability-aware UAV assignment, and PPO-based adaptive coordination, characterizing their expected behavior under dynamically changing disaster conditions without assuming unconditional convergence.

%%%%%%%%%%%%%%%%%%%%%%%%%%%%%%%%%%%%%%%%%%%%%%%%%%%%%%%%%%%%%%%%%%%%%%%%%%
\subsection{Mission Capability Satisfaction Analysis}
\label{subsec:mission_capability_satisfaction}

Let the total rescue capability allocated to adaptive rescue region
$R_j$ at time $t$ be $C_j(t)$. The corresponding capability deficit is
defined as
$ \Delta_j(t) = \left[MCR_j(t)-C_j(t)\right]^+,
[z]^+=\max(z,0).$
The \ac{CSR} is defined as
$ CSR(t) = \frac{1}{K} \sum_{j=1}^{K} \mathbb{1}
\left( \Delta_j(t)=0 \right),$
where $\mathbb{1}(\cdot)$ is the indicator function.
The average regional capability deficit is 
%%%%%%%%%%%%%%%%%%%%%%%%%%%%%%%%
\begin{equation}
\overline{\Delta}(t)
= \frac{1}{K} \sum_{j=1}^{K} \Delta_j(t).
\label{eq:average_capability_deficit}
\end{equation}
%%%%%%%%%%%%%%%%%%%%%%%%%%%%%%%%
Suppose that during one reallocation interval, a total capability
$u_j(t)\geq0$ is transferred to $R_j$. The updated regional
deficit is
%%%%%%%%%%%%%%%%%%%%%%%%%%%%%%%%%
\begin{equation}
\Delta_j(t+1) = \left[ MCR_j(t+1) - C_j(t) - u_j(t)
\right]^+.
\label{eq:updated_capability_deficit}
\end{equation}
%%%%%%%%%%%%%%%%%%%%%%%%%%%%%%%%%
For a fixed regional requirement during one scheduling interval,
i.e.,
$MCR_j(t+1)=MCR_j(t),$
the deficit evolution reduces to
$
\Delta_j(t+1) = \left[ \Delta_j(t)-u_j(t) \right]^+.$
Since $u_j(t)\geq0$, it follows directly that
$ 0 \leq \Delta_j(t+1) \leq \Delta_j(t). $
Therefore, capability reallocation cannot increase the capability deficit of the receiving region during a scheduling interval with
fixed $MCR_j$.
Furthermore, if the transferred capability satisfies $u_j(t)\geq\Delta_j(t),$ then
$\Delta_j(t+1)=0.$
Hence, with sufficient transferable capability, the receiving region's
deficit can be eliminated. For fixed regional requirements, an effective
reallocation cannot decrease the number of capability-satisfied regions $CSR(t+1)\geq CSR(t)$. However, in a dynamic disaster environment, $MCR_j(t)$ may increase due to newly detected victims, environmental degradation, or changing rescue requirements. Thus, the framework does not claim unconditional convergence
to $CSR=1$ but continuously seeks high capability satisfaction under
dynamic mission conditions.

%%%%%%%%%%%%%%%%%%%%%%%%%%%%%%%%%%%%%%%%%%%%%%%%%%%%%%%%%%%%%%%%%%%%%%%%%%
\subsection{Rescue Delay Analysis}

Consider an adaptive rescue region $R_j$ containing $V_j$
unresolved victims. Assume that the effective rescue service rate
is proportional to the total rescue capability allocated to the
region
$ \mu_j = a_r C_j, $
where $a_r>0$ is the rescue-efficiency coefficient.
Under this idealized deterministic service-rate model, the
corresponding rescue service delay is defined as
%%%%%%%%%%%%%%%%%%%%%%%%%%%%%%%%%
\begin{equation}
D_j = \frac{V_j}{\mu_j} = \frac{V_j}{a_r C_j}.
\label{eq:regional_rescue_delay}
\end{equation}
%%%%%%%%%%%%%%%%%%%%%%%%%%%%%%%%%
Substituting the allocated capability from
(\ref{eq:initial_allocated_capability}) gives
%%%%%%%%%%%%%%%%%%%%%%%%%%%%%%%%%
\begin{equation}
D_j = \frac{V_j} { a_r \displaystyle\sum_{i=1}^{N}
x_{ij}\mathrm{MCS}_i^{*} }.
\label{eq:delay_capability_relationship}
\end{equation}
%%%%%%%%%%%%%%%%%%%%%%%%%%%%%%%%%
For $C_j>0$, the first derivative of the delay with respect to
allocated capability is expressed as
$ \frac{\partial D_j}{\partial C_j} = -\frac{V_j}{a_r C_j^2} <0.$
The second derivative is
$ \frac{\partial^2 D_j}{\partial C_j^2} = \frac{2V_j}{a_r C_j^3} >0.$
Therefore, rescue delay is strictly decreasing and convex with
respect to the allocated rescue capability. This means that
increasing the capability assigned to an under served region
reduces its idealized rescue delay.
For an additional feasible capability increment
$z C_j>0$
%%%%%%%%%%%%%%%%%%%%%%%%%%%%%%%%%
\begin{equation}
D_j(C_j+z C_j)-D_j(C_j)
= -\frac{V_j z C_j} {a_r C_j(C_j+z C_j)} <0.
\label{eq:incremental_delay_reduction}
\end{equation}
%%%%%%%%%%%%%%%%%%%%%%%%%%%%%%%%%
The total idealized rescue delay over all adaptive regions is
%%%%%%%%%%%%%%%%%%%%%%%%%%%%%%%%%
\begin{equation}
D_{\mathrm{total}} = \sum_{j=1}^{K}
D_j = \sum_{j=1}^{K} \frac{V_j}{a_r C_j}.
\label{eq:total_rescue_delay}
\end{equation}
%%%%%%%%%%%%%%%%%%%%%%%%%%%%%%%%%
Hence, the analytical result directly supports the proposed
\ac{MCR}--\ac{MCS}-based capability allocation: regions with insufficient
allocated capability have greater potential rescue delay and
benefit from additional capable \ac{UAV} resources.

%%%%%%%%%%%%%%%%%%%%%%%%%%%%%%%%%%%%%%%%%%%%%%%%%%%%%%%%%%%%%%%%%%%%%%%%%%
\subsection{Energy Consumption Analysis}
For rescue \ac{UAV} $U_i$ following trajectory $\Gamma_i$, the
propulsion energy is given by
$ E_i = \int_{0}^{k_i} P(v_i(t))\,dt, $
where $P(v_i(t))$ denotes the propulsion power at velocity
$v_i(t)$. For the adopted rotary-wing \ac{UAV} model, propulsion power $P(v)$ is defined as
%%%%%%%%%%%%%%%%%%%%%%%%%%%%%%%%%
\begin{equation}
\begin{aligned}
P_0
\left(
1+\frac{3v^2}{U_{\mathrm{tip}}^2}
\right) + P_i \left( \sqrt{
1+\frac{v^4}{4v_0^4} }
- \frac{v^2}{2v_0^2} \right)^{1/2}
+ \frac{1}{2}
d_0 p_s A v^3,
\label{eq:rotary_wing_power}
\end{aligned}
\end{equation}
%%%%%%%%%%%%%%%%%%%%%%%%%%%%%%%%%
where $P_0$ and $P_i$ denote the blade-profile and induced-power
constants, respectively, $U_{\mathrm{tip}}$ is the rotor-tip
speed, $v_0$ is the induced velocity in hover, $d_0$ is the
fuselage drag ratio, $p_s$ is the air density, and $A$ is the
rotor-disc area.
The total fleet propulsion energy is therefore
$ E_{\mathrm{total}} = \sum_{i=1}^{N}E_i. $
For the discrete trajectory representation used in the proposed
optimization, the energy can be approximated by
$ E_i
\approx \sum_{\ell=0}^{L-1} P
\left( \left\| \mathbf{v}_i(\ell) \right\| \right)
\Delta t.$
Consider two feasible trajectories
$\Gamma_i^{(1)}$ and $\Gamma_i^{(2)}$ for same \ac{UAV} and
mission. Their energy difference is
$ \Delta E_i = E_i(\Gamma_i^{(1)}) -E_i(\Gamma_i^{(2)}).$
If the proposed trajectory satisfies
$ E_i(\Gamma_i^{\mathrm{prop}}) <
E_i(\Gamma_i^{\mathrm{base}}), $
then the corresponding energy reduction is
%%%%%%%%%%%%%%%%%%%%%%%%%%%%%%%%%
\begin{equation}
\eta_E = \frac{ E_i(\Gamma_i^{\mathrm{base}})
- E_i(\Gamma_i^{\mathrm{prop}}) }{ E_i(\Gamma_i^{\mathrm{base}}) } \times100\%.
\label{eq:energy_reduction_percentage}
\end{equation}
%%%%%%%%%%%%%%%%%%%%%%%%%%%%%%%%%
Thus, a lower optimized trajectory energy directly indicates improved propulsion-energy efficiency for the same rescue mission.
The simulation results subsequently quantify this theoretical
relationship by comparing the proposed trajectory optimization
against the selected baseline strategy.

%%%%%%%%%%%%%%%%%%%%%%%%%%%%%%%%%%%%%%%%%%%%%%%%%%%%%%%%%%%%%%%%%%%%%%%%%%
\subsection{System Stability Analysis}
To analyze the capability-deficit dynamics, consider the Lyapunov
candidate
%%%%%%%%%%%%%%%%%%%%%%%%%%%%%%%%%
\begin{equation}
V(t) = \frac{1}{2} \sum_{j=1}^{K} \Delta_j^2(t),
\label{eq:lyapunov_candidate}
\end{equation}
%%%%%%%%%%%%%%%%%%%%%%%%%%%%%%%%%
where $\Delta_j(t)$ denotes the capability deficit of region
$R_j$. Since $\Delta_j(t)\geq0$, it follows that
$ V(t)\geq0,$ with $V(t)=0$ if and only if all adaptive regions satisfy their
capability requirements.
Let $u_j(t)\geq0$ denote the capability transferred to region
$R_j$. For a fixed  during one scheduling interval is defined as
%%%%%%%%%%%%%%%%%%%%%%%%%%%%%%%%%
\begin{equation}
\Delta_j(t+1) = [\Delta_j(t)-u_j(t)]^+.
\label{eq:stability_deficit_dynamics}
\end{equation}
%%%%%%%%%%%%%%%%%%%%%%%%%%%%%%%%%
For a transfer satisfying
$ 0\leq u_j(t)\leq\Delta_j(t), $
the positive-part operator becomes inactive and
$ \Delta_j(t+1) = \Delta_j(t)-u_j(t). $
The Lyapunov difference $\Delta V(t)$ is therefore denoted as
%%%%%%%%%%%%%%%%%%%%%%%%%%%%%%%%%
\begin{align}
&=
V(t+1)-V(t)
\nonumber=
\frac{1}{2}
\sum_{j=1}^{K}
\left[
(\Delta_j(t)-u_j(t))^2
-
\Delta_j^2(t)
\right]
\nonumber\\
&=
-\sum_{j=1}^{K}
\Delta_j(t)u_j(t)
+
\frac{1}{2}
\sum_{j=1}^{K}
u_j^2(t).
\label{eq:lyapunov_difference}
\end{align}
%%%%%%%%%%%%%%%%%%%%%%%%%%%%%%%%%
Equivalently, $\Delta V(t)
= -\sum_{j=1}^{K} u_j(t) \left( \Delta_j(t) \frac{u_j(t)}{2} \right).$
Therefore, if
$ 0<u_j(t)<2\Delta_j(t) $
for the actively served deficient regions, then
$ \Delta V(t)<0.$
Hence, under fixed regional requirements and bounded effective
capability transfers, the aggregate squared capability deficit
decreases after an effective reallocation step.
When sufficient transferable capability exists to eliminate all
regional deficits, $V(t)=0
\Longleftrightarrow
\Delta_j(t)=0,
\ \forall j.$ Thus, the capability-deficit dynamics exhibit a local Lyapunov
stability property between successive mission updates. Because
the disaster environment is time-varying, $MCR_j(t)$ may change
between scheduling intervals. Therefore, this result should not
be interpreted as an unconditional global stability or convergence
guarantee.

%%%%%%%%%%%%%%%%%%%%%%%%%%%%%%%%%%%%%%%%%%%%%%%%%%%%%%%%%%%%%%%%%%%%%%%%%%

\subsection{Computational Complexity Analysis}
The proposed optimization jointly determines the binary \ac{UAV} assignment matrix and continuous trajectories, resulting in a mixed-integer nonlinear programming problem. The assignment matrix $X=[x_{ij}]$ contains $NK$ binary variables, while the discretized trajectory set
$ \Gamma=\{\mathbf{p}_1(t),\mathbf{p}_2(t),\ldots,\mathbf{p}_N(t)\} $
introduces approximately $O(NL)$ continuous position variables for $L$ decision instants. Thus, the decision variables scale as $O(NK+NL)$. With fixed trajectories, the capability-aware assignment remains an NP-hard generalized assignment-type problem. Exhaustive enumeration has a search space of $O(K^N)$; for $N=50$ and $K=10$, this yields $10^{50}$ possible assignments, making exhaustive real-time optimization impractical. This motivates the proposed \ac{DRL}-based optimization mechanism for obtaining high-quality assignment and trajectory decisions efficiently.

%%%%%%%%%%%%%%%%%%%%%%%%%%%%%%%%%%%%%%%%%%%%%%%%%%%%%%%%%%%%%%%%%%%%%%%%%%
\section{Simulation Results} \label{smi}
The proposed framework is evaluated through simulations under varying
disaster conditions, rescue demands, and \ac{UAV} operational states,
and compared with greedy, distance-based, battery-aware, and random
baselines. The system comprises $N=50$ rescue \acp{UAV} and $K=10$
adaptive regions over a $1000\times1000$~m$^2$ area. The rescue \acp{UAV} operate at altitudes of
 $100$~m, respectively, with $V_{\max}=20$~m/s,
$P_t=1$~W, and $B=20$~MHz. The victim population varies from
$100$ to $700$. The sensing range and local \ac{UAV}
communication range are set to $250$~m and $300$~m,
respectively, while master-platform connectivity is evaluated
using the \ac{HAPS}-to-\ac{UAV} channel model. The \ac{AI} confidence
threshold is set to $c_{\mathrm{th}}=0.90$.
\subsection{Communication and Mission-Level Performance}

Fig.~\ref{fig:performance_comparison}(a) evaluates the proposed capability-aware coordination framework communication reliability against greedy, battery-aware, distance-based, and random assignment techniques.
The proposed framework achieves approximately $99\%$ communication
reliability, compared with approximately $95\%$, $92\%$, $89\%$, and
$83\%$ for the greedy, battery-aware, distance-based, and random
strategies, respectively. This corresponds to improvements of
approximately $4.2\%$ over the greedy strategy and $19.3\%$ over the
random strategy. The improvement results from explicitly incorporating
communication quality into the mission-capability evaluation and
UAV-region association. Consequently, \acp{UAV} with limited
communication capability, they are less likely to be assigned to regions with
high communication requirements, enabling more reliable
master-\ac{HAPS}--rescue-\ac{UAV} connectivity during mission execution.

Fig.~\ref{fig:performance_comparison}(b) presents the total flight distance as the number of victims increases from $100$ to $700$. The flight distance increases for all approaches as the growing victim population increases rescue demand and requires additional \ac{UAV} movements. However, the proposed framework consistently achieves the lowest flight distance. At $700$ victims, it requires approximately $990$~m, compared with $1120$~m, $1300$~m, $1240$~m, and $1510$~m for the greedy, distance-based, battery-aware, and random strategies, respectively. This corresponds to reductions of approximately $11.6\%$, $23.8\%$, $20.2\%$, and $34.4\%$, respectively. These results indicate that capability-aware coordination selects \acp{UAV} according to their suitability for the assigned rescue region rather than relying solely on spatial proximity or battery status, thereby reducing unnecessary \ac{UAV} movements and inefficient region-to-\ac{UAV} associations.

Fig.~\ref{fig:performance_comparison}(c) shows the mission completion time as the victim count increases. The proposed framework consistently achieves the shortest completion time. At $700$ victims, it completes the mission in approximately $60$~min, compared with $72$, $90$, $84$, and $105$~min for the greedy, distance-based, battery-aware, and random strategies, respectively. This corresponds to reductions of approximately $16.7\%$, $33.3\%$, $28.6\%$, and $42.9\%$, respectively. The improvement is mainly attributed to the joint consideration of regional \ac{MCR} and \ac{UAV} \ac{MCS}, which enables suitable \acp{UAV} to serve demanding regions, reducing capability mismatches and repeated reassignment. Thus, the capability-aware coordination mechanism becomes increasingly beneficial as the rescue workload increases.

\subsection{Rescue Efficiency and Energy Performance}
\begin{figure*}[t]
\centering
\subfloat[]{
  \includegraphics[width=56mm,height=40mm]{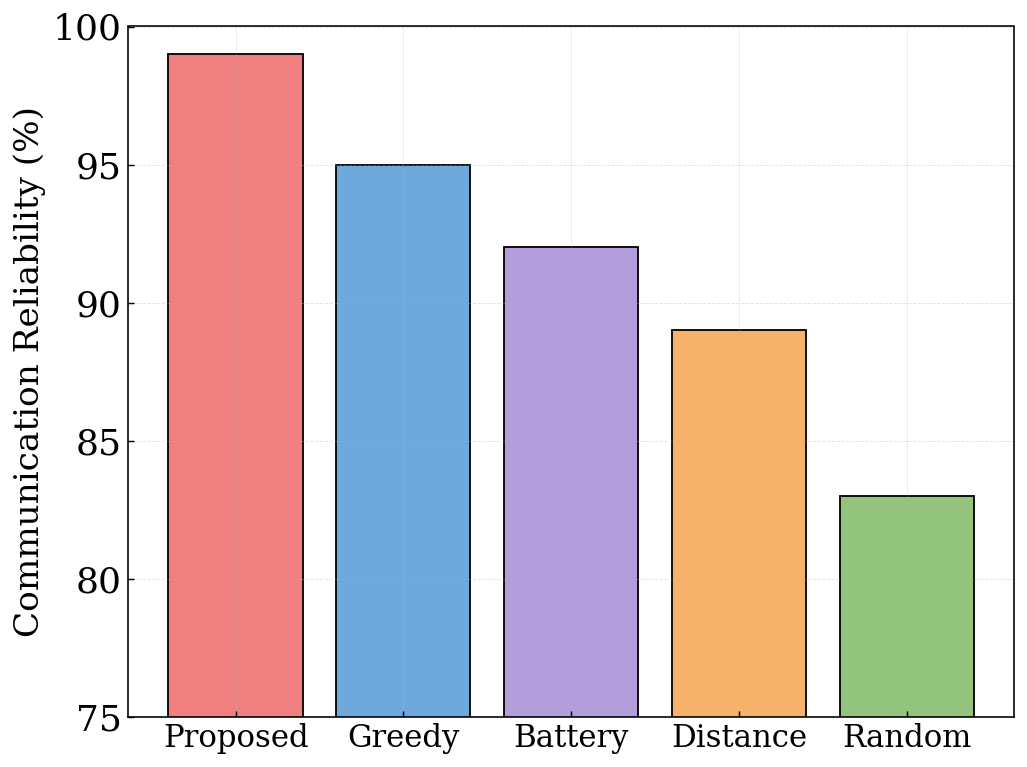}
}
\subfloat[]{
  \includegraphics[width=56mm,height=40mm]{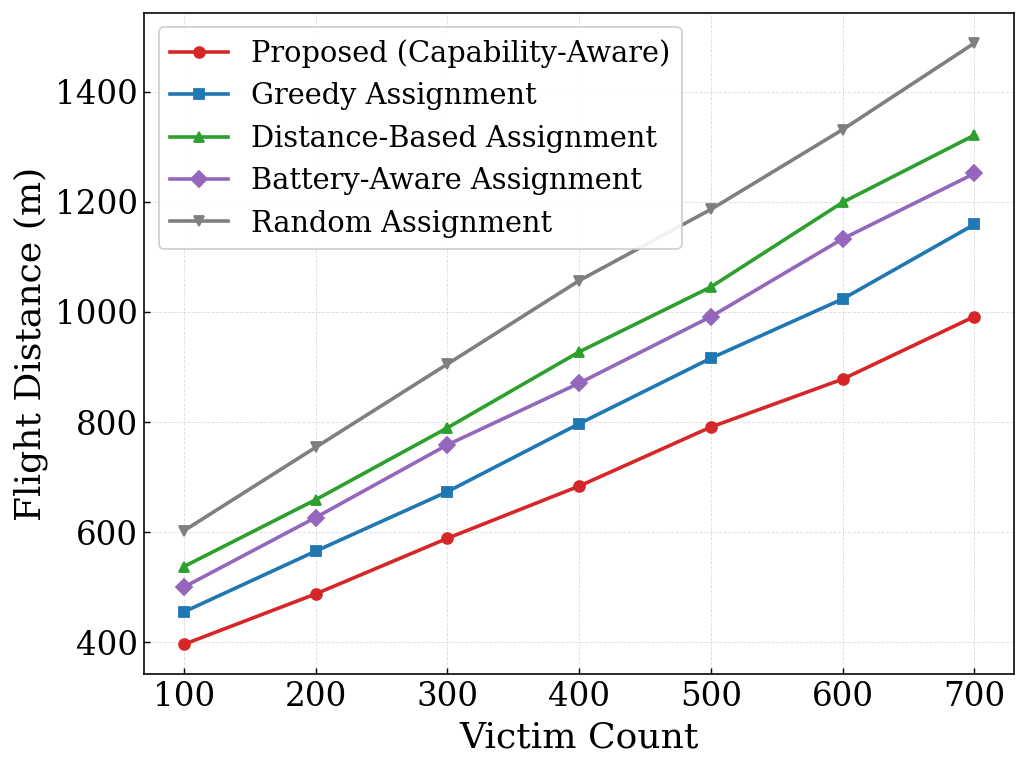}
}
\subfloat[]{
  \includegraphics[width=56mm,height=40mm]{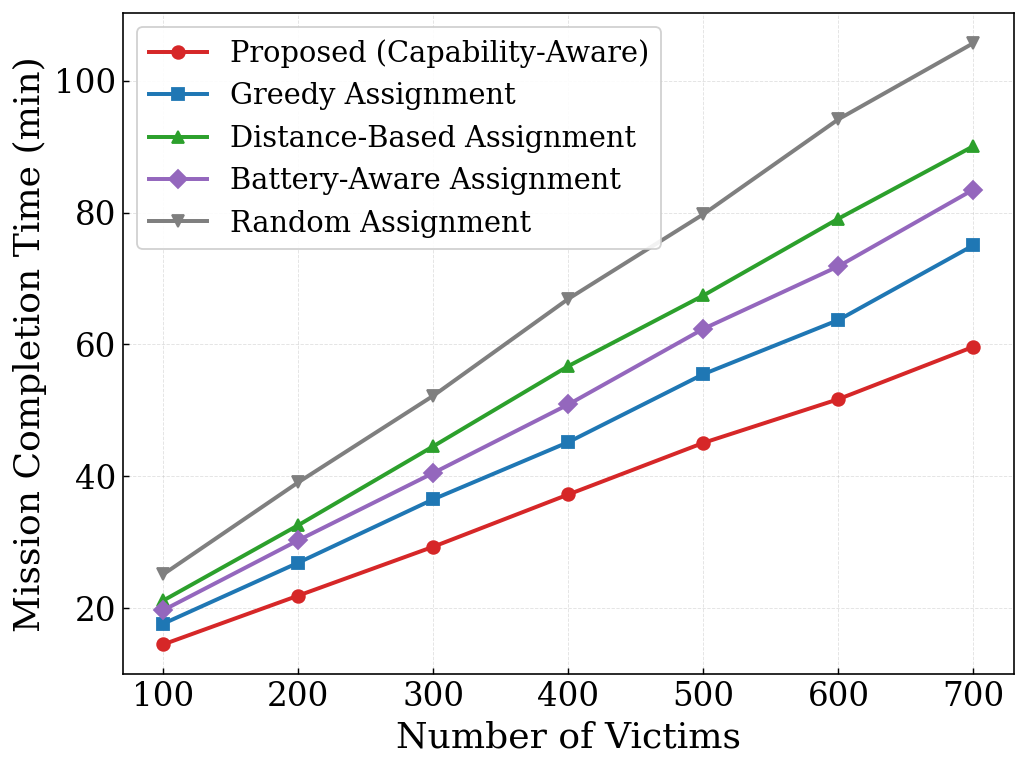}
}

\subfloat[]{
  \includegraphics[width=56mm,height=40mm]{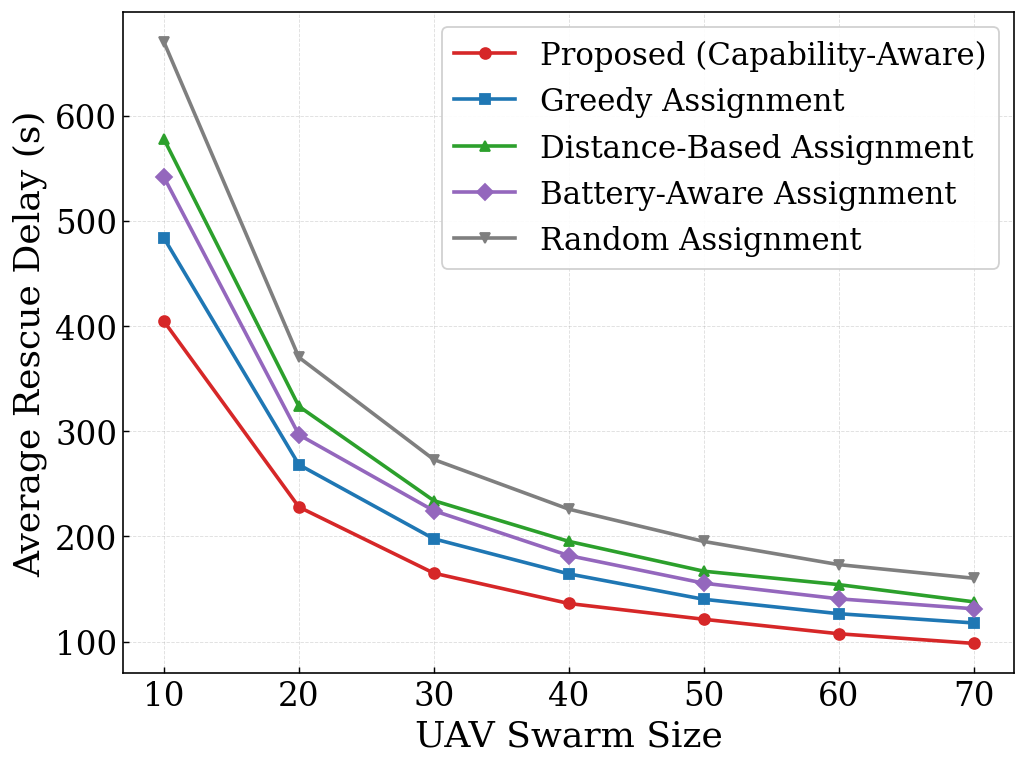}
}
\subfloat[]{
  \includegraphics[width=56mm,height=40mm]{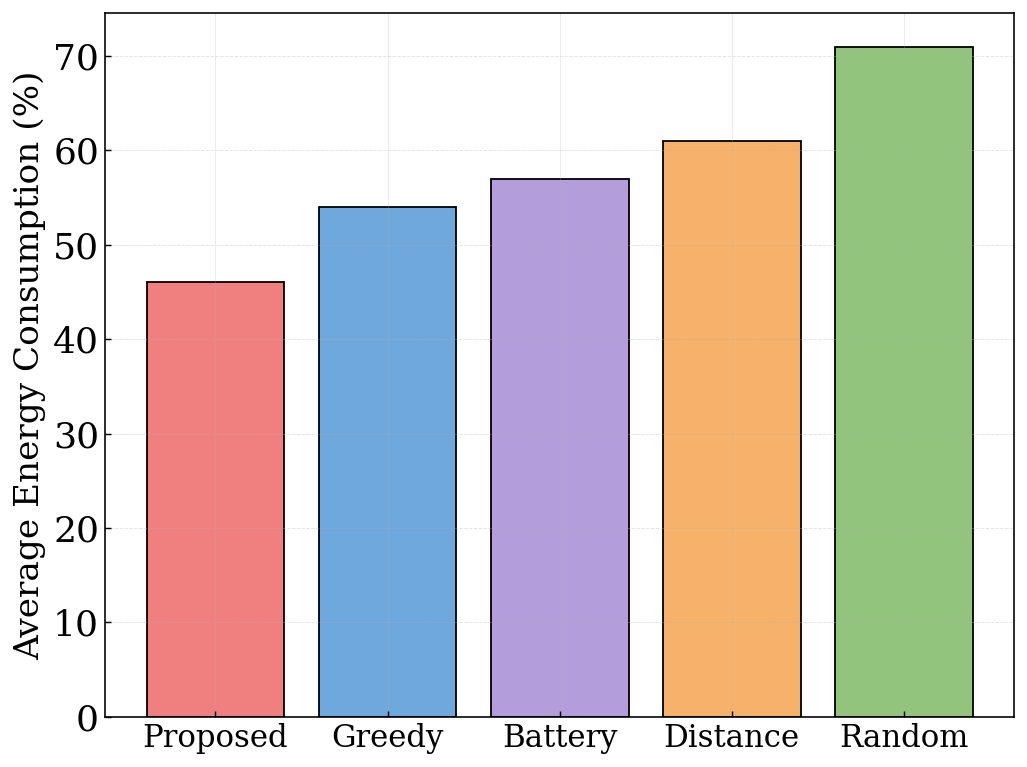}
}
\subfloat[]{
  \includegraphics[width=56mm,height=40mm]{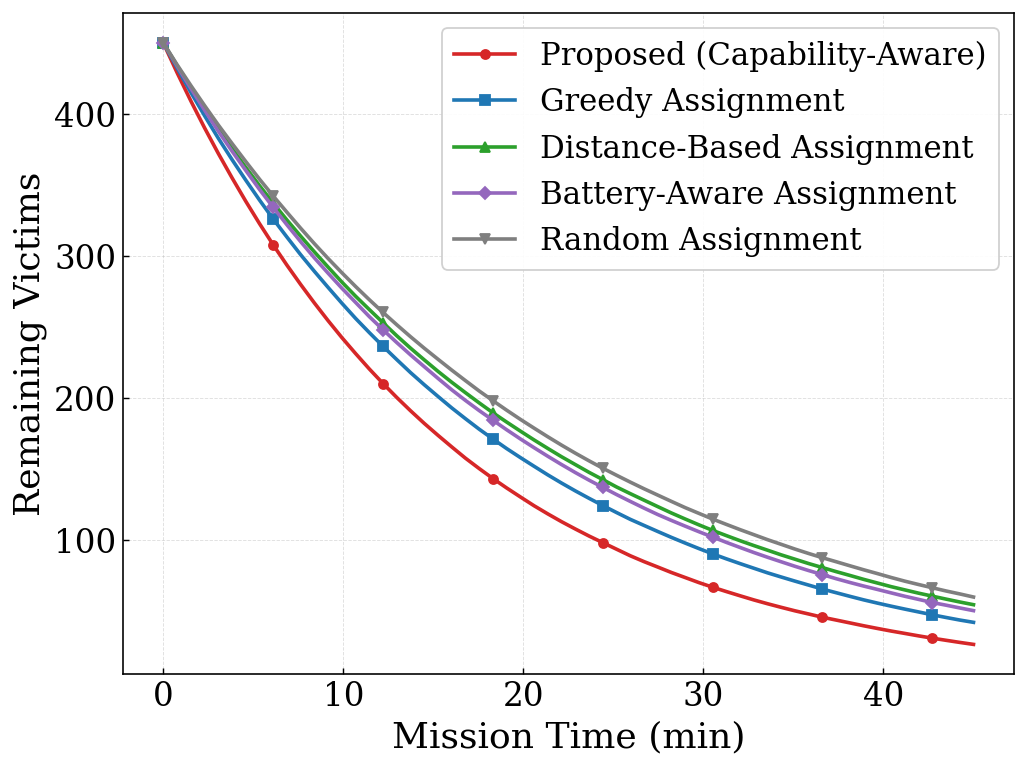}
}

\caption{Performance comparison of the proposed and baseline strategies in terms of (a) communication reliability, (b) total flight distance, (c) mission completion time, (d) average rescue delay, (e) energy consumption, and (f) remaining victims.}
\label{fig:performance_comparison}

\end{figure*}

Fig.~\ref{fig:performance_comparison}(d) shows the average rescue delay time for various swarm sizes of \acp{UAV}. Because more rescue \acp{UAV} can concurrently serve various adaptive rescue regions, the average rescue delay for all studied techniques lowers as the number of rescue \acp{UAV} grows. Across all swarm sizes taken into consideration, the proposed capability-aware method consistently yields the lowest delay. This behavior results from matching the operational capability
of an individual \acp{UAV} with the capability requirements of their assigned
regions. Consequently, highly demanding regions are less likely to
remain under served, reducing the waiting time of victims. The
decreasing trend further demonstrates that the proposed framework can
effectively exploit additional \ac{UAV} resources without relying solely on
geographical proximity.

\begin{figure}[t]
\centering
\subfloat[]{
  \includegraphics[width=41.9mm,height=37mm]{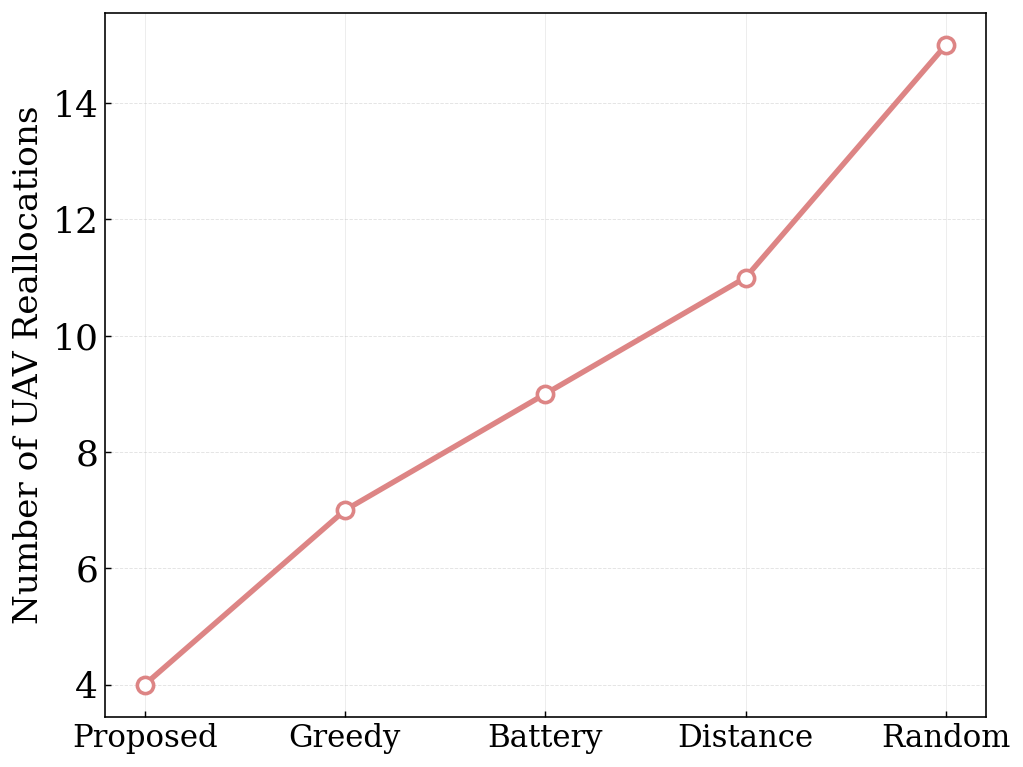}
}
\subfloat[]{
  \includegraphics[width=41.9mm,height=37mm]{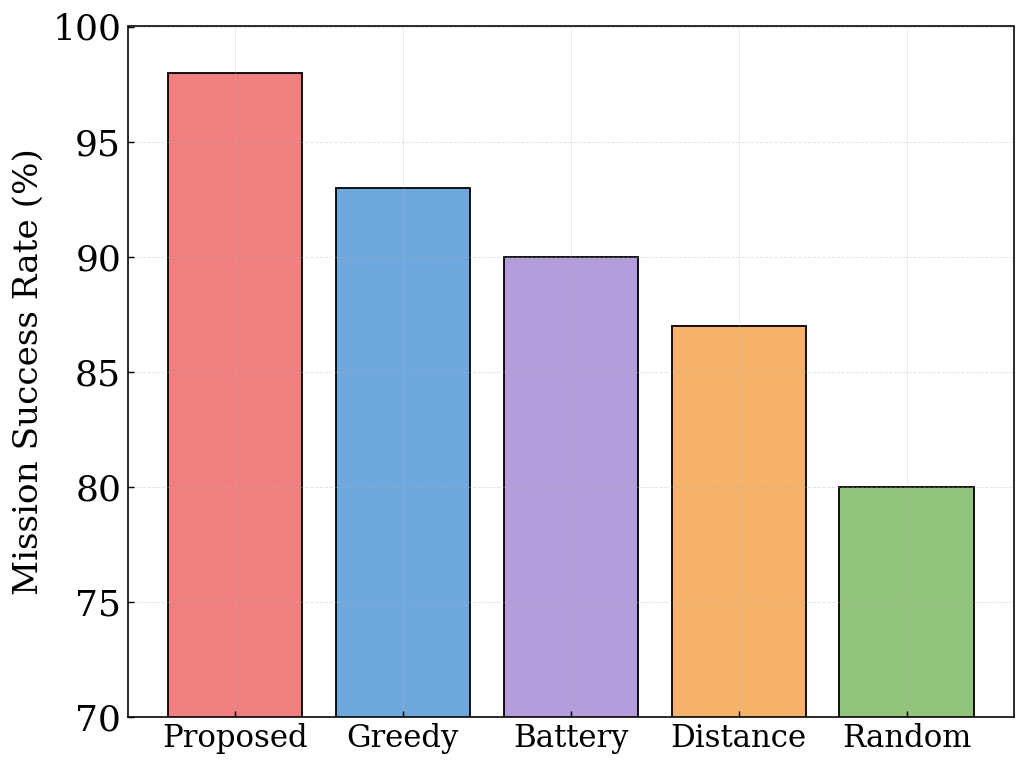}
  
}

\subfloat[]{
  \includegraphics[width=41.9mm,height=37mm]{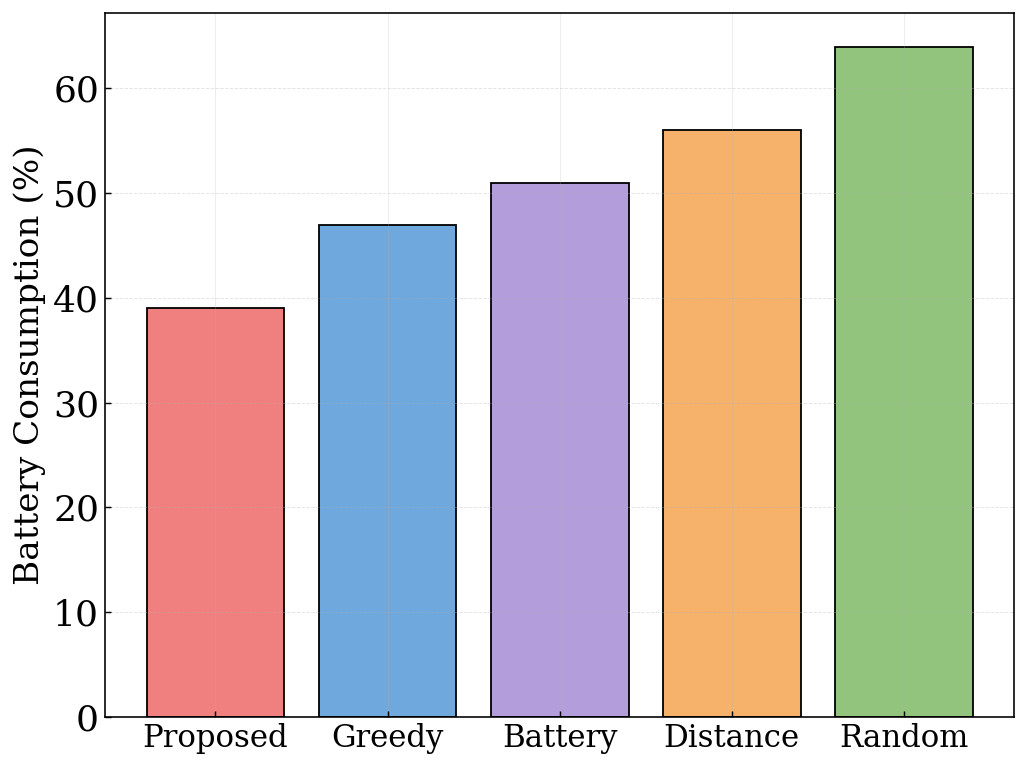}
}
\subfloat[]{
  \includegraphics[width=41.9mm,height=37mm]{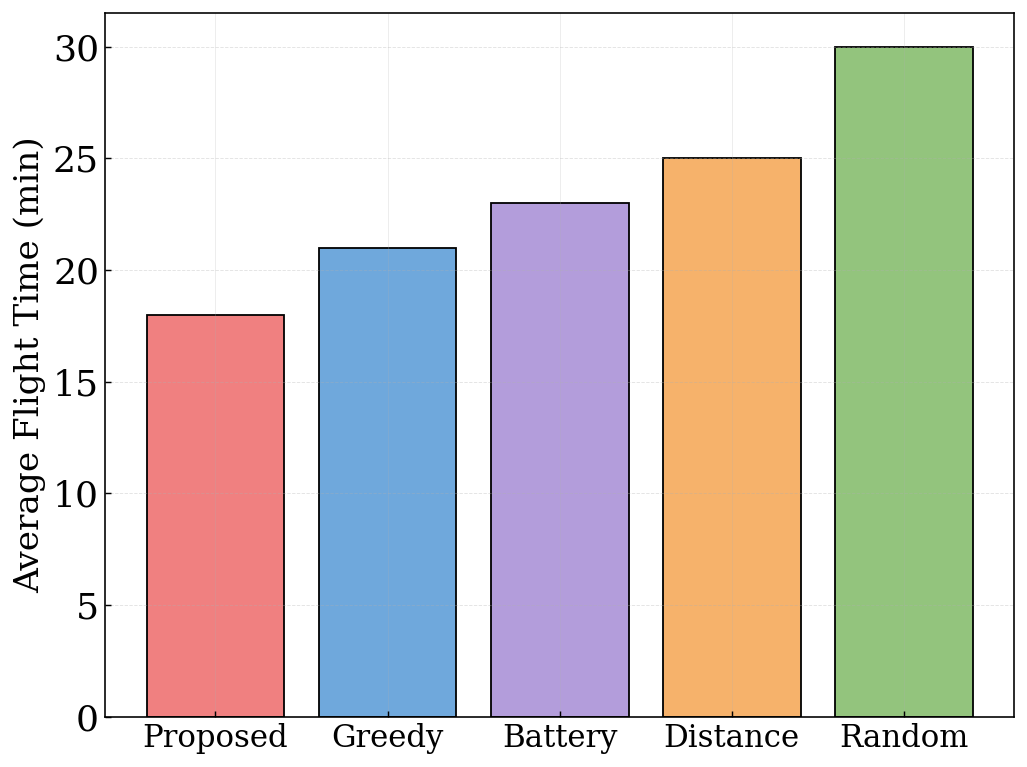}
}

\subfloat[]{
  \includegraphics[width=41.9mm,height=37mm]{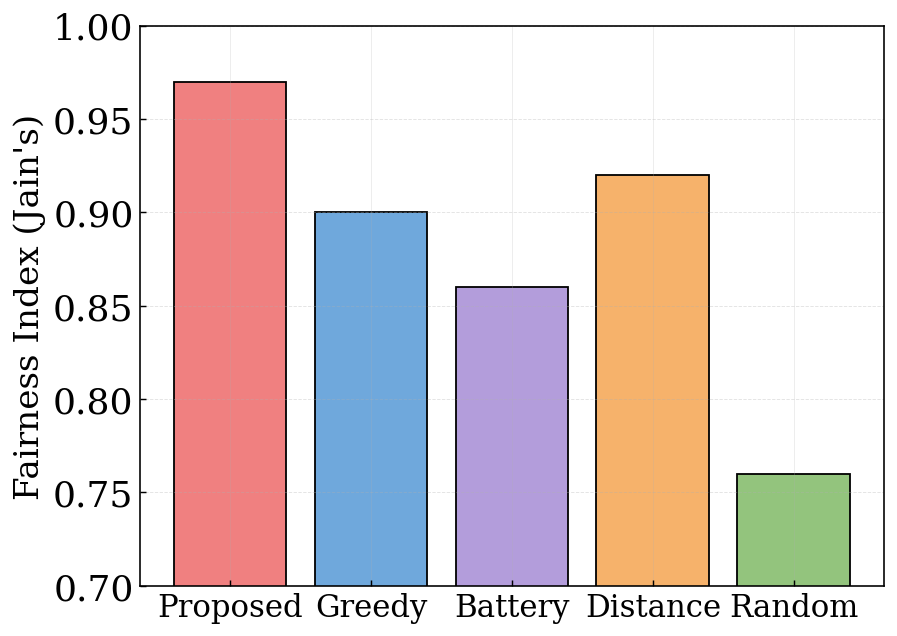}
}
\subfloat[]{
  \includegraphics[width=41.9mm,height=37mm]{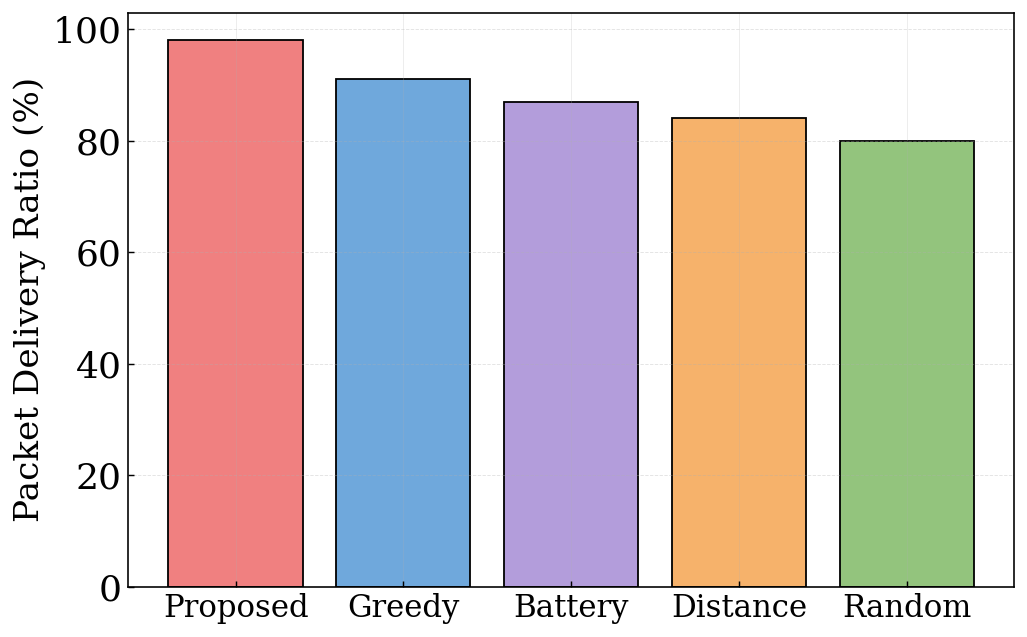}
}
\caption{(a) \ac{UAV} reallocation, (b) mission success rate, (c) battery consumption, (d) average flight time, (e) Jain's fairness index, and (f) packet delivery ratio.}
\label{fig:performance_comparison uav}

\end{figure}

Fig.~\ref{fig:performance_comparison}(e) compares the average energy
consumption of the evaluated strategies. The proposed framework
achieves the lowest energy consumption throughout the considered
scenarios. The reduction is attributed to capability-aware \ac{UAV}
assignment and dynamic reallocation, which avoid repeatedly
deploying \acp{UAV} that are poorly suited to the requirements of a region.
By selecting \acp{UAV} with appropriate residual battery, sensing,
communication, endurance, and workload conditions, the proposed
framework reduces inefficient movement and unnecessary resource
usage. This result is particularly important in disaster-response
missions because excessive energy consumption can cause them \acp{UAV} to leave
the operational area before completing their assigned rescue tasks.

Fig.~\ref{fig:performance_comparison}(f) shows the number of remaining victims versus mission time. The number of unresolved victims decreases for all approaches as rescue operations progress. The proposed AI-driven capability-aware framework achieves a faster reduction by dynamically prioritizing regions based on evolving rescue requirements and assigning suitable \ac{UAV} capabilities. This adaptive coordination accelerates victim servicing and reduces overall rescue delay.

\subsection{Dynamic Reallocation and Mission Robustness}

Fig.~\ref{fig:performance_comparison uav}(a) evaluates the quantity of \ac{UAV} reallocation needed by the strategies under consideration. In comparison to the greedy, battery-aware, distance-based, and random techniques, which require $7$, $9$, $11$, and $15$ reallocation, the proposed framework requires about $4$. This corresponds to reductions of approximately $43\%$,$55.6\%$, $63.6\%$, and $73.3\%$, respectively. The lower reallocation frequency indicates that capability-aware initial assignment produces more suitable \ac{UAV}--region associations. Furthermore, the \ac{MCR}--\ac{MCS} mechanism triggers \ac{UAV} transfer only when the existing resources cannot satisfy a capability deficit, thereby avoiding unnecessary migration and reducing coordination overhead.

Fig.~\ref{fig:performance_comparison uav}(b) presents the mission
success rate of the considered allocation strategies. The proposed
framework achieves a mission success rate of  $\approx 98\%$,
compared with approximately $93\%$, $90\%$, $87\%$, and $80\%$ for
the greedy, battery-aware, distance-based, and random strategies,
respectively. This corresponds to relative improvements of
approximately $5.4\%$, $8.9\%$, $12.6\%$, and $22.5\%$ over the
respective baselines. The improvement demonstrates that satisfying
regional capability requirements is more effective than considering
individual \ac{UAV} parameters independently. In particular, the \ac{MCR}--\ac{MCS}
matching mechanism reduces the probability that a \ac{UAV} with insufficient
operational capability is assigned to a demanding rescue region.

Fig.~\ref{fig:performance_comparison uav}(c) illustrates the
percentage of the available battery budget consumed during the rescue
mission. The proposed framework consumes  $\approx39\%$ of the
available battery budget, compared with approximately $47\%$, $51\%$,
$56\%$, and $64\%$ for the greedy, battery-aware, distance-based, and
random strategies, respectively. Therefore, the proposed framework
reduces battery consumption by approximately $17.0\%$, $23.5\%$,
$30.4\%$, and $39.1\%$, respectively. The lower battery consumption
is achieved through more suitable \ac{UAV} assignment, reduced unnecessary
travel, and capability-aware reallocation. These characteristics
increase the available operational margin of the \ac{UAV} swarm and can
potentially improve mission robustness under prolonged disaster
conditions.
%%%%%%%%%%%%%%%%%%%%%%%%%%%%%%%%%%%%%%%%%%%%%%%%%%%%%%%%%%%%%%%%%%%%%%%%%%
\subsection{Workload Fairness and Network Reliability}

Fig.~\ref{fig:performance_comparison uav}(d) compares the average flight time
of the evaluated strategies. The proposed framework achieves an
average flight time of approximately $18$~min, compared with
approximately $21$, $23$, $25$, and $30$~min for the greedy,
battery-aware, distance-based, and random strategies, respectively.
This corresponds to reductions of approximately $14.3\%$, $21.7\%$,
$28.0\%$, and $40.0\%$ relative to the corresponding baselines.
The lower flight time indicates that capability-aware assignment
reduces inefficient \ac{UAV} movement while maintaining effective rescue
coverage. This behavior is consistent with the lower total flight
distance observed in Fig.~\ref{fig:performance_comparison}(b).

Fig.~\ref{fig:performance_comparison uav}(e) evaluates workload
distribution using Jain's Fairness Index. The proposed framework
achieves fairness index of $\approx 0.97$, compared with
approximately $0.92$, $0.90$, and $0.76$ for the distance-based,
greedy, and random strategies, respectively. The high fairness value
indicates that the proposed capability-aware mechanism distributes
rescue workload more uniformly among the \acp{UAV}. This is important
because repeatedly assigning demanding regions to a small subset of
\acp{UAV} may rapidly deplete their battery and reduce their remaining
mission capability. By incorporating workload and operational
capability into \ac{UAV} selection, the proposed framework maintains a more
balanced swarm workload.

Fig.~\ref{fig:performance_comparison uav}(f) presents the packet delivery
ratio (PDR) achieved by the considered strategies. The proposed
framework achieves a PDR of approximately $98\%$, outperforming the
greedy, battery-aware, distance-based, and random strategies, which
achieve approximately $91\%$, $87\%$, $84\%$, and $80\%$, respectively.
The proposed approach therefore improves PDR by approximately
$7.7\%$, $12.6\%$, $16.7\%$, and $22.5\%$ relative to the respective
baselines. The improvement indicates that maintaining suitable
communication capability during \ac{UAV} assignment and trajectory
execution contributes to more reliable information exchange between
the master \ac{HAPS} and rescue \acp{UAV}.

%%%%%%%%%%%%%%%%%%%%%%%%%%%%%%%%%%%%%%%%%%%%%%%%%%%%%%%%%%%%%%%%%%%%%%%%%%
\section{Conclusion}
\label{con}

This paper presented an \ac{AI}-driven mission capability-aware
\ac{UAV} swarm coordination framework for autonomous post-disaster
\ac{SAR}. The framework integrates \ac{ISAC} sensing, confidence-aware
\ac{AI} perception, adaptive disaster partitioning,
\ac{MCR}--\ac{MCS} capability matching, dynamic \ac{UAV} reallocation,
and hierarchical \ac{DRL} coordination to adapt resources to evolving
regional rescue requirements. Simulation results demonstrate improved
capability satisfaction, reduced rescue delay, and more efficient
resource utilization compared with the considered baselines. Future work will focus on adaptive mission-capability management under
dynamic disaster conditions and an \ac{UAV} resource allocation.

%%%%%%%%%%%%%%%%%%%%%%%%%%%%%%%%%%%%%%%%%%%%%%%%%%%%%%%%%%%%%%%%%%%%%%%%%%

\bibliographystyle{IEEEtran}
\bibliography{references}
\end{document}